\documentclass[aps,prl,reprint,groupedaddress,showpacs]{revtex4-1}
\usepackage{amsmath}
\usepackage{amssymb}
\usepackage{graphicx}
\usepackage{array}
\usepackage{dcolumn}
\usepackage{subfigure}
\usepackage{color}
\usepackage{float}
\usepackage[hidelinks=true]{hyperref}
\hypersetup{
  colorlinks   = true, 
  urlcolor     = blue, 
  linkcolor    = blue, 
  citecolor   = red 
}

\begin{document}
\title{An atomistic model of electronic polarizability for calculation of Raman scattering from large-scale MD simulations}

\author{Atanu Paul$^1$}
\author{Anthony Ruffino$^3$}
\author{Stefan Masiuk$^2$}
\author{Jonathan Spanier$^{2,3}$}
\author{Ilya Grinberg$^1$}
\email[]{ilya.grinberg@biu.ac.il}

\affiliation{$^1$Department of Chemistry, Bar-Ilan University, Ramat Gan 5290002, Israel}

\affiliation{$^2$Department of Mechanical Engineering and Mechanics, Drexel University, Philadelphia, Pennsylvania 19104, USA}

\affiliation{$^3$Department of Physics, Drexel University, Philadelphia, Pennsylvania 19104, USA}

\begin{abstract}
The application of molecular dynamics (MD) simulations to the interpretation  of Raman scattering spectra  is hindered by inability of atomistic simulations to account for the dynamic evolution of electronic polarizability, requiring the use of either $ab$ $initio$ method or parameterization of machine learning models.  
More broadly, the dynamic evolution of electronic-structure-derived properties cannot be treated by the current atomistic models.  
Here, we report a simple, physically-based atomistic model with few (maximum 10 parameters for the systems considered here) adjustable parameters that can accurately represent the changes in the electronic polarizability tensor for molecules and solid-state systems. Due to its compactness, the model can be applied for simulations of Raman spectra of large ($\sim 1,000,000$-atom) systems with modest computational cost.  To demonstrate its accuracy, the model is applied to the CO$_2$ molecule, water clusters, and  BaTiO$_{3}$ and CsPbBr$_{3}$
perovskites and shows good agreement with $ab$-$initio$-derived and experimental polarizability tensor and Raman data. The atomistic nature of the model  enables  local analysis of the contributions to Raman spectra,  paving the way for the application of MD simulations for the interpretation of Raman spectroscopy results.  Furthermore, our successful atomistic representation of the evolution of  electronic polarizability suggests that the evolution of  electronic structure and its derivative properties  can be represented by atomistic models, opening up the possibility of studies of electronic-structure-dependent properties using large-scale atomistic  simulations.

\end{abstract}

\maketitle

Raman scattering spectroscopy is  a powerful tool for the study of structure and dynamics of solid, liquid and gas-phase materials~\cite{Butler2016, Raman_review}. 
Similar to infrared (IR) spectra, the interpretation of Raman spectra can benefit strongly from the use of molecular dynamics (MD) simulations that allow the decomposition of the total Raman spectrum into the contributions of individual modes or structural features~\cite{CPB_PRL}. However, since the Raman spectrum must be obtained from MD simulations using the Fourier transform of the electronic polarizability tensor time autocorrelation function~\cite{Raman_intensity}, the derivation of Raman spectra from MD simulations requires the calculation of electronic polarizability for each time step in the MD simulation. While electronic polarizability can be calculated using  
density functional perturbation theory (DFPT)~\cite{DFPT_baroni}, such calculations are 
suitable only for simulations of small systems and short simulation times.

Several approaches have been used to circumvent this difficulty by creating models for estimating the electronic polarizability without full $ab$ $initio$ calculations. The bond-polarizability model expresses the changes in the polarizability contribution of a bond as a function of the bond lengths assuming the absence of interactions between the bonds and has been successfully used to reproduce Raman spectra of selected molecules and nanotubes~\cite{PhysRevB.53.13106,PhysRevB.71.241402, ghosez}. However, this method is challenging to use in solids, where multi-atom bonding is important. Recently, an efficient method has been proposed to calculate spectra for large solids with impurities and alloys~\cite{HannuPekka, HannuPekka1} based on the projection of dynamics on the Raman active modes. While applicable to large systems, this method is based on the first-order Raman scattering and therefore considers phonons around the $\Gamma$ point only, which is incomplete for systems such as Si, SrTiO$_{3}$, and MoS$_{2}$~\cite{Si_Raman,STO_Raman,Livneh_2015}.  
\begin{figure*}[]
\centering 
\includegraphics[width = 17.6 cm,angle =0]{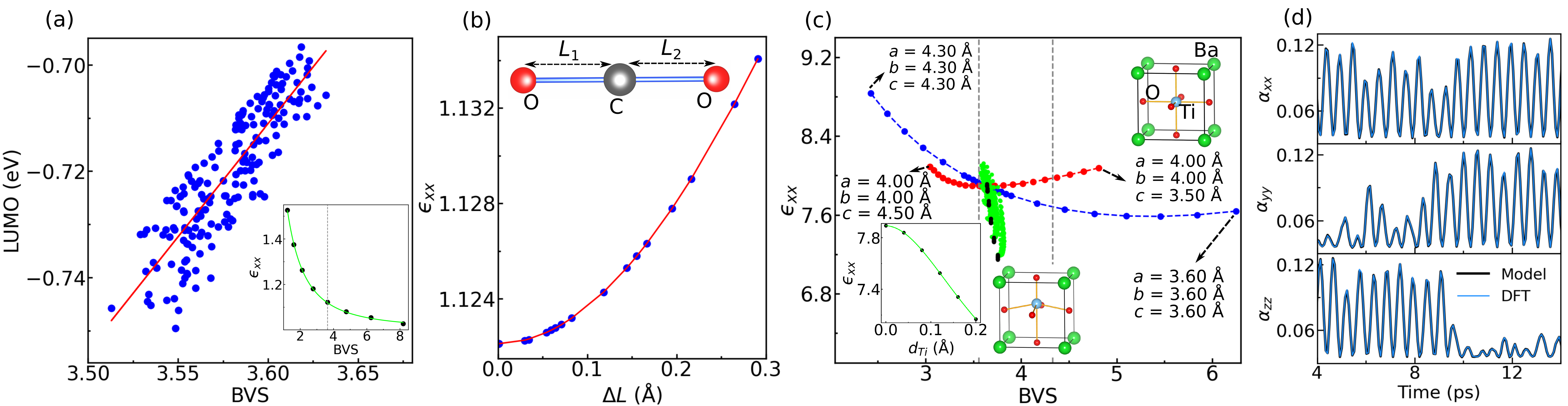}
\caption{(Color online) (a) LUMO (eV) vs BVS (blue circles) and the linear fit (red line) for  CO$_{2}$ structures generated by MD at 300 K. Inset: the Lorentzian function fit (green) of $\epsilon_{xx}$ to  BVS (black dots) for symmetrically stretched CO$_{2}$. Gray dotted line (Inset): BVS (3.64) of CO$_{2}$ at DFT-relaxed position. (b) $\epsilon_{xx}$ vs $\Delta L$ ($\mid L_{1}$ - $L_{2}\mid$) for CO$_{2}$ for BVS = 3.64 (blue circles) with the Lorentzian fit (red). $L_{1}$ and $L_{2}$ are C-O bond lengths (see inset). (c) $\epsilon_{xx}$ vs Ti BVS for five-atom BTO for isotropic volume change of cubic centrosymmetric BTO (blue), compression and extension of the $z$-axis  of centrosymmetric BTO (red), Ba and Ti shifted along $x$-direction while O are fixed at centrosymmetric positions for cubic BTO with $a$ = 4.00 \AA~(black circles), Ba and Ti fixed at the centrosymmetric positions while O are shifted along $x$-direction for cubic BTO with $a$ = 4.00 \AA~(green circles). Gray dotted lines: Minimum (3.56) and maximum (4.33) Ti BVS from 5-atom MD in the temperature range 200-1000 K.  
Bottom corner inset: Lorentzian fit (green) of $\epsilon_{xx}$ vs off-centre displacement of Ti ($d_{Ti}$ (\AA)) (black dots), where Ba and O are at centrosymmetric position. 
(d) Polarizability ($\alpha_{xx}$, $\alpha_{yy}$ and $\alpha_{zz}$) trajectory using model and DFT for CO$_{2}$ at 300 K.}
\label{Fig1}
\end{figure*}

Alternatively, the polarizability surface (i.e. polarizability as a function of atomic coordinates) can be modeled using either polynomial expansion or deep learning methods to enable direct calculation of  Raman spectra from MD trajectories~\cite{polynomial,car,analytica3030020}. However, both of these approaches require a large database of training structure for the parameterization of the model. Deep learning models are also 
more computationally expensive (by a factor of up to 100) than atomistic potentials 
~\cite{nonVonNeumann,ShiLiu,PhysRevLett.120.143001,9355242}.  
However, atomistic models for polarizability similar in spirit to the atomistic models for evaluation of energy and forces from atomic coordinates have not been reported to date for complex systems with multi-center bonding. This is may be due to the assumption that the response of electronic structure to the incident electromagnetic field and its dependence on atomic coordinates in a solid-state  material are too complicated to be decomposed into contributions of individual atoms and expressed by a simple atomistic model and thus can only be modeled using quantum mechanical approaches.  

Here, we demonstrate that  by considering the  effect of the dynamical changes in the structure and chemical bonding in molecules or materials together with the  second-order perturbation theory expression for  electronic polarizability, a computationally efficient and atomistically-interpretable model with few adjustable parameters characterizing the evolution of electronic polarizability in terms of atomic coordinates  can be derived. 
We then apply the model with $\leq$ 10 parameters for the calculation of Raman spectra from MD simulations using CO$_{2}$, water clusters ((H$_{2}$O)$_{n}$: $n$ = 1-6, 8), and BaTiO$_3$ (BTO) and CsPbBr$_3$ (CPB) 
systems with supercells of up to $\approx$ 500,000 atoms  to demonstrate the accuracy of this approach.  

From perturbation theory, the polarizability tensor (\pmb{$\alpha$}) can be represented as~\cite{hutter,long} 
$$
\alpha_{ij} \propto \sum_{\gamma\neq 0} \frac{\langle\Phi_{0}\mid\hat{p_{i}}\mid\Phi_{\gamma} \rangle \langle\Phi_{\gamma}\mid\hat{p_{j}}\mid\Phi_{0}\rangle}{\hbar(\omega_{\gamma} - \omega_{0})}
$$
where $\mid\Phi_{0} \rangle$ and $\mid\Phi_{\gamma} \rangle$ are ground and excited states, $\hat{p_{i}}$ is the induced electric dipole operator due to external field, and $\hbar$($\omega_{\gamma} - \omega_{0}$) is the energy difference between excited and ground states. 
Thus, the atomistic model  must take into account the effect of the changes of the bonding and the  coupling between the individual bonds on this perturbation theory expression. The denominator ($\omega_{\gamma} - \omega_{0}$) can be related to the second moment of the local density of states and thus the bond valence sum (BVS) ~\cite{Sutton,Liu13} defined as
\begin{equation}
\textnormal{BVS} = \sum_{i} exp((d^{0}_{AB} - d_{i})/b)
\label{eq2}
\end{equation}
where $d^{0}_{AB}$ and $d_{i}$ are the bond valence parameter and bond length for the bond between atoms $A$ and $B$, respectively, $b$ (= 0.37 \AA) is an universal constant~\cite{BVS}.
Therefore, BVS rather than the individual bond lengths characterizes the changes in the local bonding environment and
should be related to the \pmb{$\alpha$}.  We use the CO$_2$ molecule as a simple model system to demonstrate the relationship between BVS and the LUMO energy (see Fig.~\ref{Fig1} (a)). Plotting $\pmb{\epsilon}$ versus BVS for a symmetric stretching of CO$_{2}$, we find that the atom's polarizabilty ($\pmb{\alpha}$) due to symmetrically stretched bond can be characterized by Lorentzian function of the BVS of that atom (see inset of Fig.~\ref{Fig1} (a))\footnote{$\alpha_{xx}$ = normalization factor$\times$($\epsilon_{xx}$ - 1) and the normalization factor is related to the amount of vacuum in the unit cell used to calculate $\epsilon_{xx}$. Both of these give similar expressions. Therefore, while plotting we have shown DFT-calculated $\epsilon_{xx}$. However, in text we always use $\alpha_{xx}$.}.

Since BVS is insensitive to the details of the local atomic arrangement, the effect of local asymmetry on $\pmb{\alpha}$ must be taken into account separately.  We investigate the influence of the asymmetry on $\pmb{\alpha}$ by plotting $\epsilon_{xx}$ versus the bond length difference ($\Delta L$)  in CO$_2$ for a fixed BVS (3.64) so that the values of ($\omega_{\gamma}$ - $\omega_{0}$) show only small variation  (Fig.~\ref{Fig1} (b)).  It is observed that $\pmb{\alpha}$ shows a Lorentzian dependence on the asymmetry of CO$_2$.

To understand the relationship between $\pmb{\alpha}$ and atomic coordinates, BVS and asymmetry in a more complicated solid-state system, we considered the following cases for a 5-atom BTO cell: (a) isotropic volume change of cubic centrosymmetric BTO, (b) Ba and Ti displacing along $x$ direction and O fixed at centrosymmetric position for cubic BTO where lattice constant = 4.00 {\AA} and (c) Ba and Ti fixed at the centrosymmetric positions and O displacing along $x$ direction for cubic BTO where lattice constant = 4.00 \AA, (d) Anisotropic volume change (tetragonal distortion) of centrosymmetric BTO. 
The plots of $\pmb{\alpha}$ versus Ti BVS for cases (a)-(d) are shown in Fig.~\ref{Fig1} (c). For (a), we  observed a Lorentzian dependence of $\pmb{\alpha}$ on BVS in the physically relevant range of $\approx$ 3.5-4.5.  For (b) and (c), despite the small variation in BVS, a strong change in $\pmb{\alpha}$ is observed with $\pmb{\alpha}$ decreasing strongly with greater off-center atomic displacement (asymmetry). The dependence of $\pmb{\alpha}$ on asymmetry (off-center displacement of Ti) also follows a Lorentzian dependence (see inset of Fig.~\ref{Fig1} (c)).  For (d), we observe an interesting trend of $\pmb{\alpha}$ first decreasing and then increasing with the change in the BTO aspect ratio. 
The results for cases (a)-(c) show that similar to the CO$_{2}$ molecule, $\pmb{\alpha}$ for the solid-state BTO shows Lorentzian dependence on the BVS  and the asymmetric atomic displacement.  
The results for (d) can be explained by considering  that more isolated atoms have higher $\pmb{\alpha}$ than the corresponding more  bonded system (in part due to lower ($\omega_{\gamma} - \omega_{0}$)) and that BTO with either very low or very high aspect ratio will contain O atoms (axial O atoms for  high aspect ratio, equatorial O atoms for low aspect ratio) that participate only weakly in bonds with Ti. For high aspect ratio, the long distances between Ti and axial O atoms will make the O atoms more non-bonding and will increase $\pmb{\alpha}$. For low aspect ratio, the very short distances between Ti and axial O atoms will lead to the concentration of the bonding along the $z$-axis, leaving the O atoms in the $xy$ plane with weak bonds to Ti even for the Ti-O distances of 2.0  {\AA}.  To represent this effect, we introduce dynamical charge $m$ for evaluation of the actual bond order of a given Ti-O bond taking into account the competition for bonding charge with all other bond. 

Based on the insights obtained from the relationships described above, we define 
the bond polarizability $\alpha_{ij}$ between atoms $A$ and $B$ where sum over bond polarizability gives total $\pmb{\alpha}$, as  
\begin{equation}
\begin{split}
\alpha_{ij} = [A_{\parallel}^{s}*A_{\parallel}^{a} + B_{\parallel}^{s}*B_{\parallel}^{a}]*(\frac{R_{i}R_{j}}{R^{2}}) \\
+ [A_{\perp}^{s}*A_{\perp}^{a} + B_{\perp}^{s}*B_{\perp}^{a}]*(\delta_{ij} -  \frac{R_{i}R_{j}}{R^{2}}) 
\end{split}
\label{eq1}
\end{equation}
where dielectric constant ($\epsilon_{ij}$) is related to $\alpha_{ij}$ as, $\alpha_{ij}$ $\propto$ ($\epsilon_{ij}$ - $\delta_{ij}$). Here $i$, $j$ denote $x$, $y$ or $z$ axes. The first and second expression of $\alpha_{ij}$ give the parallel and perpendicular contribution of the bond to $\alpha_{ij}$, respectively. $A_{\parallel}^{s}$ ($B_{\parallel}^{s}$) and $A_{\parallel}^{a}$ ($B_{\parallel}^{a}$)  describe the effects of the symmetric  and asymmetric bond length changes on the contribution of $A$ ($B$)  ion to the polarizability along the $A-B$ bond direction.
Similarly, $A_{\perp}^{s}$ ($B_{\perp}^{s}$)  and $A_{\perp}^{a}$ ($B_{\perp}^{a}$) describe the effects of the symmetric  and asymmetric bond length changes on the contribution of $A$ ($B$)  ion to the polarizability perpendicular to  the $A-B$ bond direction.
Here $\textbf{\textit{R}}$ defines the $A-B$ bond length.  
$\delta_{ij}$ = 1 if $i$ = $j$ and $\delta_{ij}$ = 0 if $i$ $\neq$ $j$, following the approach developed for the bond polarizability model~\cite{ghosez}.

We define $A_{\parallel}^{s}$, $A_{\parallel}^{a}$, $B_{\parallel}^{s}$, $B_{\parallel}^{a}$, $A_{\perp}^{s}$, $A_{\perp}^{a}$, $B_{\perp}^{s}$ and $B_{\perp}^{a}$ using a simple Lorentzian function. Thus we have the following expressions,\\ 
$A_{\parallel}^{s} = \frac{a_{1}^2}{(x_{A}^{s} - a_{2})^2 + a_{3}^2}$, $B_{\parallel}^{s} = \frac{a_{1}^2}{(x_{B}^{s} - a_{2})^2 + a_{3}^2}$, $A_{\perp}^{a}$ = $\frac{a_{8}^2}{(y_{A,\perp}^{a})^2 + a_{8}^2}$\\
$A_{\perp}^{s} = \frac{a_{4}^2}{(x_{A}^{s} - a_{5})^2 + a_{6}^2}$, $B_{\perp}^{s} = \frac{a_{4}^2}{(x_{B}^{s} - a_{5})^2 + a_{6}^2}$, $B_{\perp}^{a}$ = $\frac{a_{10}^2}{(y_{B,\perp}^{a})^2 + a_{10}^2}$\\
$A_{\parallel}^{a}$ = $\frac{a_{7}^2}{(y_{A,\parallel}^{a})^2 + a_{7}^2}$, $B_{\parallel}^{a}$ = $\frac{a_{9}^2}{(y_{B,\parallel}^{a})^2 + a_{9}^2}$

In the above expressions $a_{1}$, $a_{2}$, $a_{3}$, $a_{4}$, $a_{5}$, $a_{6}$, $a_{7}$, $a_{8}$, $a_{9}$, and $a_{10}$, are the constants in the Lorentzian expressions to be fit to DFPT-calculated polarizability data.  $x_{A}^{s}$, $x_{B}^{s}$, $y_{A}^{a}$ and $y_{B}^{a}$ are determined from the atomic coordinates.  $x_{A}^{s}$ is defined as,
$x_{A}^{s} = BVS\times\frac{m}{m_{0}}$
here $BVS$ = total bond valence of $A$ given by Eq.~\ref{eq2},  $m$ is the dynamical formal charge of $A$ ion that varies  depending on the environment, and  $m_{0}$ is the formal charge of $A$ (e.g. 4 for C and Ti in CO$_2$ and BTO and 2 for O). The dynamical charge state is defined as, $m$ = $\sum_{k} [\frac{BV_{k}}{BVS_{k}}\times m_{0k}]$. Here, $k$ is the neighbor bond index of $A$. $BV_{k}$ is the bond valence of the $k$-th $A$-$B$ bond. $BVS_{k}$ and $m_{0k}$ are the total bond valence and formal charge of $B$ ion residing at $k$-th $A$-$B$ bond, respectively. $x_{s,B}$ is defined in similar manner to  $x_{s,A}$.

Then, $y_{A}^{a}$ is the asymmetric displacement of $A$ ion from the center of mass of the nearest-neighbor  $B$ ions. $y_{A,\parallel}^{a}$ and $y_{A,\perp}^{a}$ are the parallel and perpendicular projection of the asymmetric displacement on the $A$-$B$ bond. $y_{B}^{a}$ is the asymmetric displacement of $B$ ion from the center of mass of the nearest-neighbor $A$ ions. $y_{B,\parallel}^{a}$ and $y_{B,\perp}^{a}$ are defined similarly to $y_{A,\parallel}^{a}$ and $y_{A,\perp}^{a}$.

To calculate these unknown parameters of Lorentzian expressions, we first calculated polarizability trajectory (\pmb{$\alpha_{DFT}$}) from density functional theory (DFT) for a small number of points along the MD trajectory. We then optimized the parameters in the  our model for $\alpha_{ij}$ (Eq.~\ref{eq1}) to reproduce the  \pmb{$\alpha_{DFT}$} using simulated annealing algorithm, with the optimized parameters given in SM. A comparison between the DFT \pmb{$\alpha$} trajectories and the model \pmb{$\alpha$} trajectories for CO$_2$ are presented  in Fig.~\ref{Fig1} (d), showing excellent agreement.  

\begin{figure}[]
\includegraphics[width = 8.5 cm,angle =0]{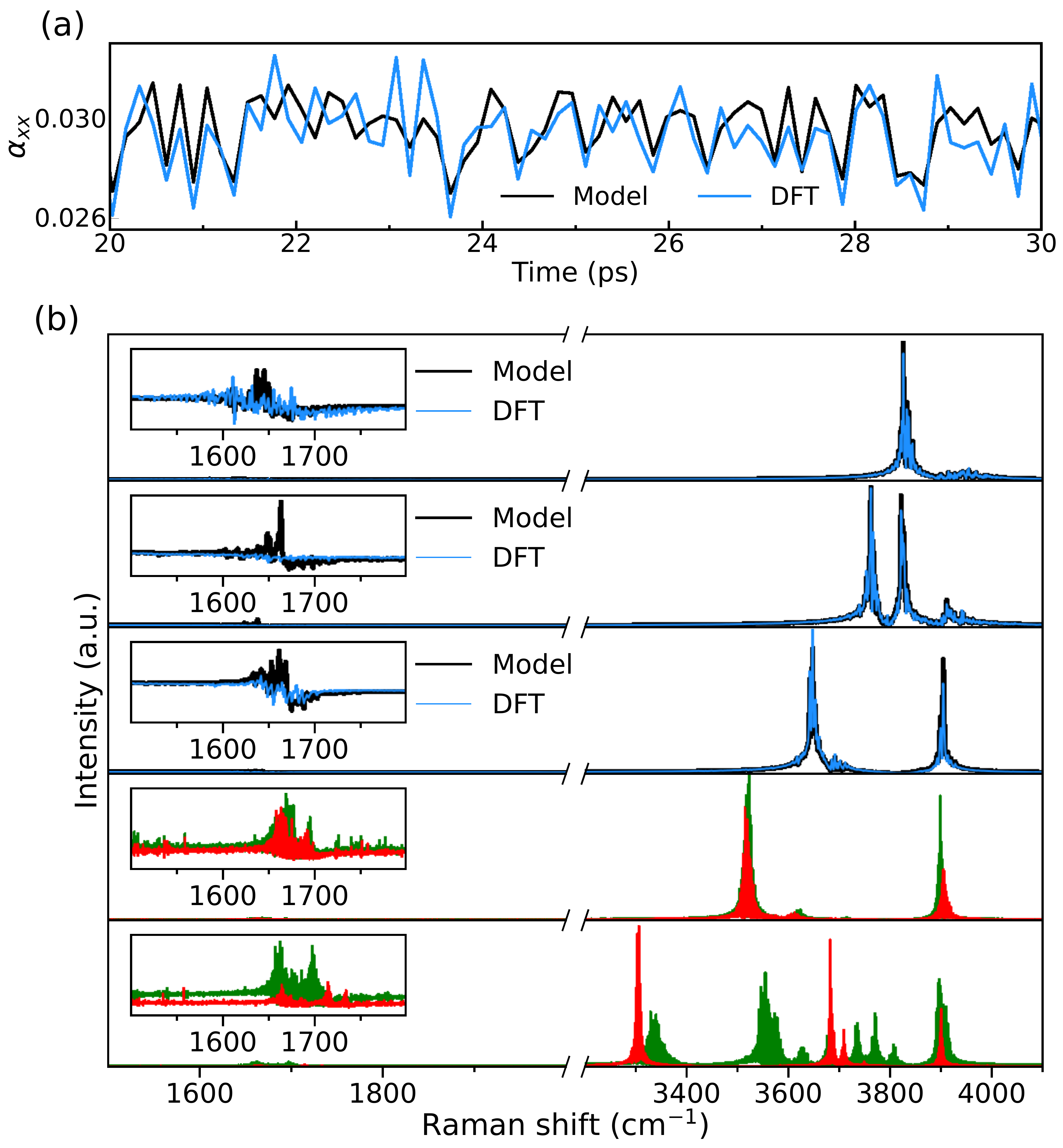}
\centering
 \caption{(Color online) (a) Polarizability ($\alpha_{xx}$) trajectory from model and DFT for H$_{2}$O molecule at 300 K. (b) Raman spectra of H$_{2}$O clusters ((H$_{2}$O)$_{n}$: $n$ = 1-6, 8). First, second and third panels show spectra of H$_{2}$O, (H$_{2}$O)$_{2}$ and (H$_{2}$O)$_{3}$ from model and DFT respectively. Fourth panel: model spectra for (H$_{2}$O)$_{4}$ (green) and (H$_{2}$O)$_{5}$ (red). Fifth panel: model spectra for (H$_{2}$O)$_{6}$ (green) and (H$_{2}$O)$_{8}$ (red).  
 }
 \label{Fig2}
\end{figure}

Next, we apply our model to calculate Raman spectra of water clusters ((H$_{2}$O)$_{n}$: $n$ = 1-6, 8). 
Fig.~\ref{Fig2} (a) show the comparison of polarizability ($\alpha_{xx}$) trajectory at 300 K as calculated from DFT and the model for H$_{2}$O. Clearly, the model trajectory captured the main features of the DFT trajectory.  We then calculate the Raman spectra of H$_{2}$O from the atomic coordinates obtained from MD simulations, 
with the results shown in Fig.~\ref{Fig2} (b)(First panel). The spectrum calculated from the DFT-obtained polarizability time autocorrelation function as shown in the same plot gives a good agreement between them. The spectra consist of three peaks, corresponding to  the H-O-H bending mode (the small peak around 1600 cm$^{-1}$), and the symmetric and asymmetric O-H  stretching modes at around 3800 cm$^{-1}$ and 3900 cm$^{-1}$ respectively. These results are in agreement  with available experimental Raman spectra ~\cite{H2O_Raman_EXPT}. 
\begin{figure}[]
\centering
\includegraphics[width = 8.6 cm,angle =0]{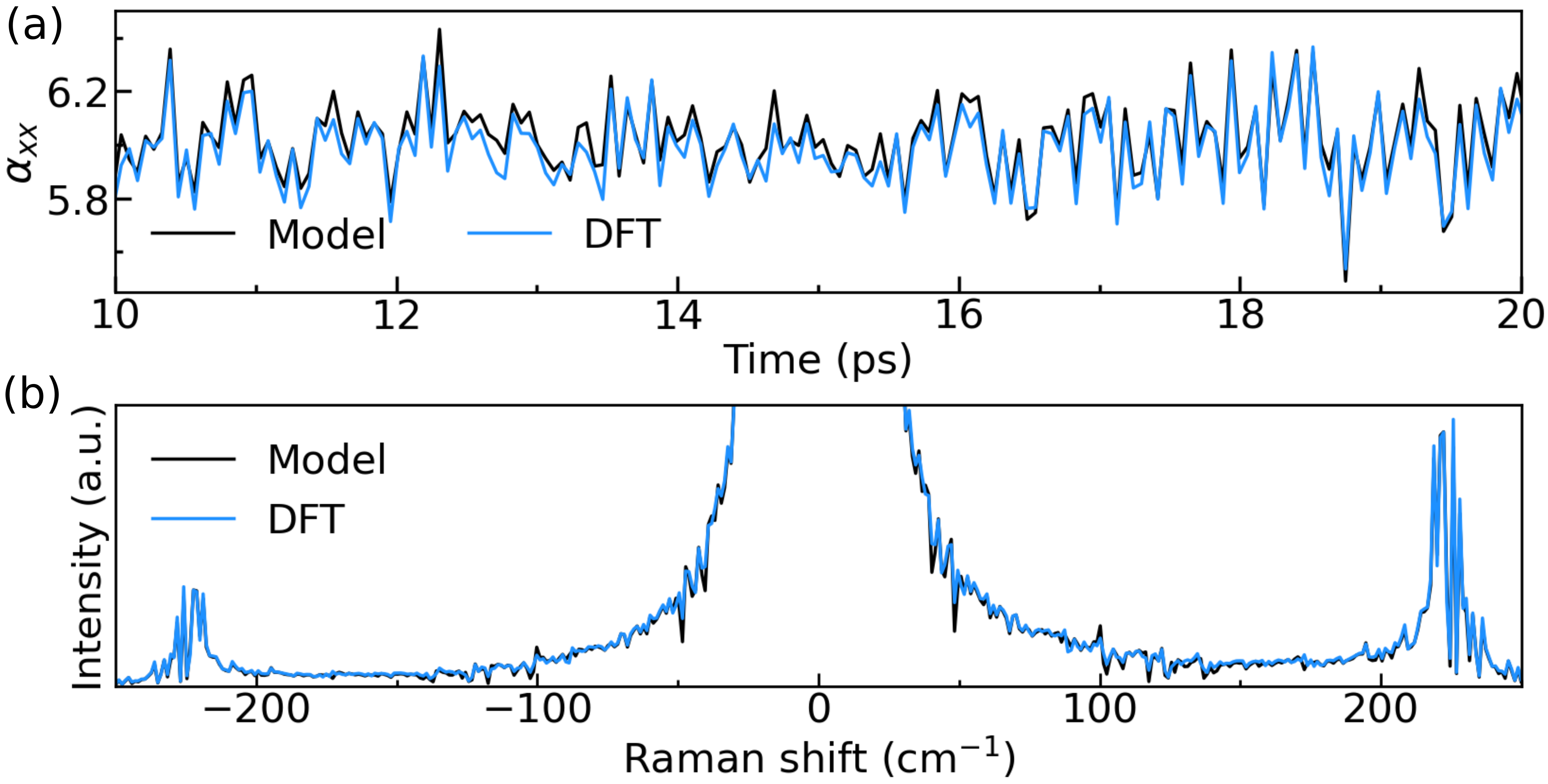}
\caption{(Color online) (a) Polarizability ($\alpha_{xx}$) trajectory, (b) Raman spectra from model and DFT for BaTiO$_{3}$ 
(5 atoms cell) at 300 K. 
}
\label{Fig3}
\end{figure}
\begin{figure*}[]
\centering
\includegraphics[width = 17.2 cm,angle =0]{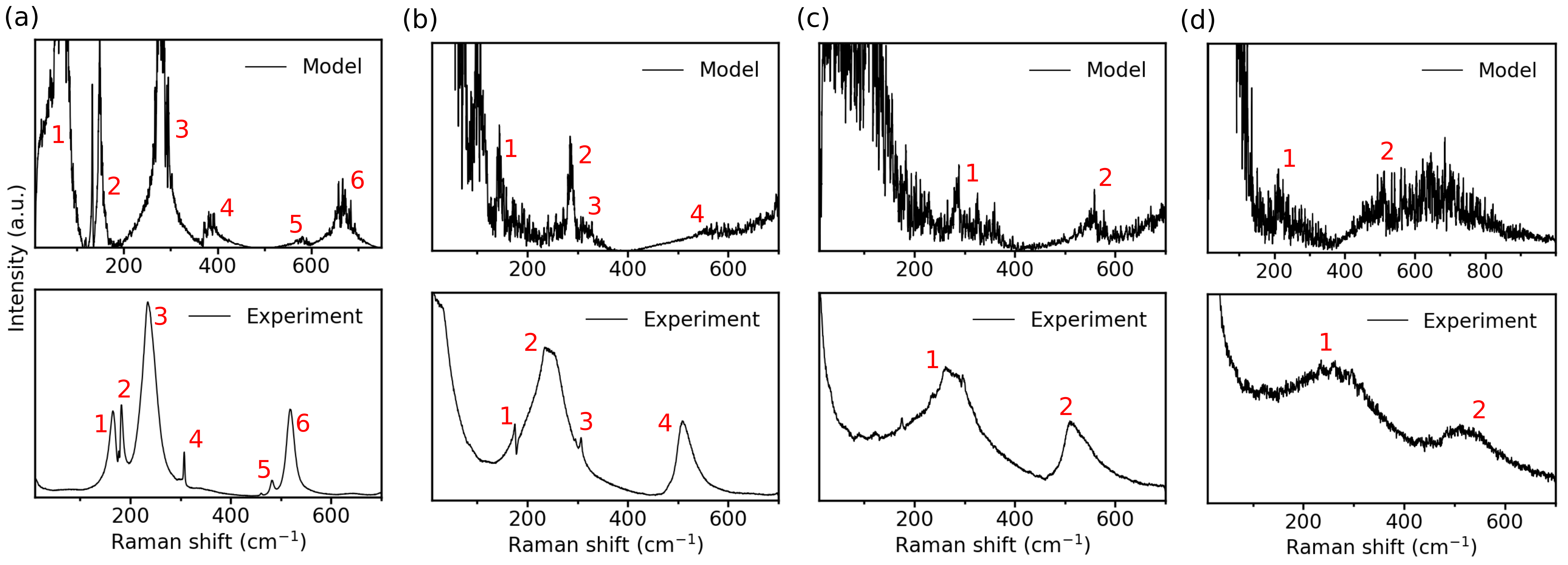}
\caption{(Color online) Raman spectra of single crystal BTO for different phases (a) rhombohedral, (b) orthorhombic, (c) tetragonal, and (d) cubic from model and experiment. The temperatures for (a), (b), (c) and (d) are 40, 92, 110 and 160 K respectively (top panel), and 123, 233, 333 and 423 K respectively (bottom panel). 
}
\label{Fig4}
\end{figure*}
To examine whether our model can be used to capture the effects of hydrogen bonding on Raman spectra, we compared the DFT and model Raman spectra of (H$_{2}$O)$_{2}$ and (H$_{2}$O)$_{3}$ clusters as calculated using the same constants as those used for single H$_{2}$O  (see Fig.~\ref{Fig2}(b)), which also show a good agreement between model and DFT results.
This shows that  the model can handle systems with hydrogen-bonded water because the effect of hydrogen bonding on polarizability is mostly due to the changes of  the H$_2$O geometry induced by hydrogen bonds, whereas hydrogen bonds induce only small changes in the response of the electron cloud to electric field for a given H$_2$O geometry.
Due to the computational constraints, the above-discussed  DFT-derived  spectra were obtained for runs of 44 ps with a time step of 0.003 ps and the same simulation duration and time step and trajectories were also used for the calculation of model-derived spectra. 
In the SM,  we present  the more accurate model Raman spectra of H$_{2}$O, (H$_{2}$O)$_{2}$ and (H$_{2}$O)$_{3}$ derived from atomistic (\cite{MBpol1, MBpol2,Naga_H2O}) simulation trajectories for 3 ns with a time step of 0.001 ps. We have also calculated the Raman spectra for (H$_{2}$O)$_{4}$, (H$_{2}$O)$_{5}$, (H$_{2}$O)$_{6}$ and (H$_{2}$O)$_{8}$ using the longer atomistic potential trajectories as shown in the
fourth and fifth panels of Fig.~\ref{Fig2}(b), respectively. The peak positions of the ((H$_{2}$O)$_{n}$, $n$ = 1-6, 8) are in good  agreement with the results in the literature~\cite{Naga_H2O}. 
\begin{figure}[]
\centering
\includegraphics[width = 8.3 cm,angle =0]{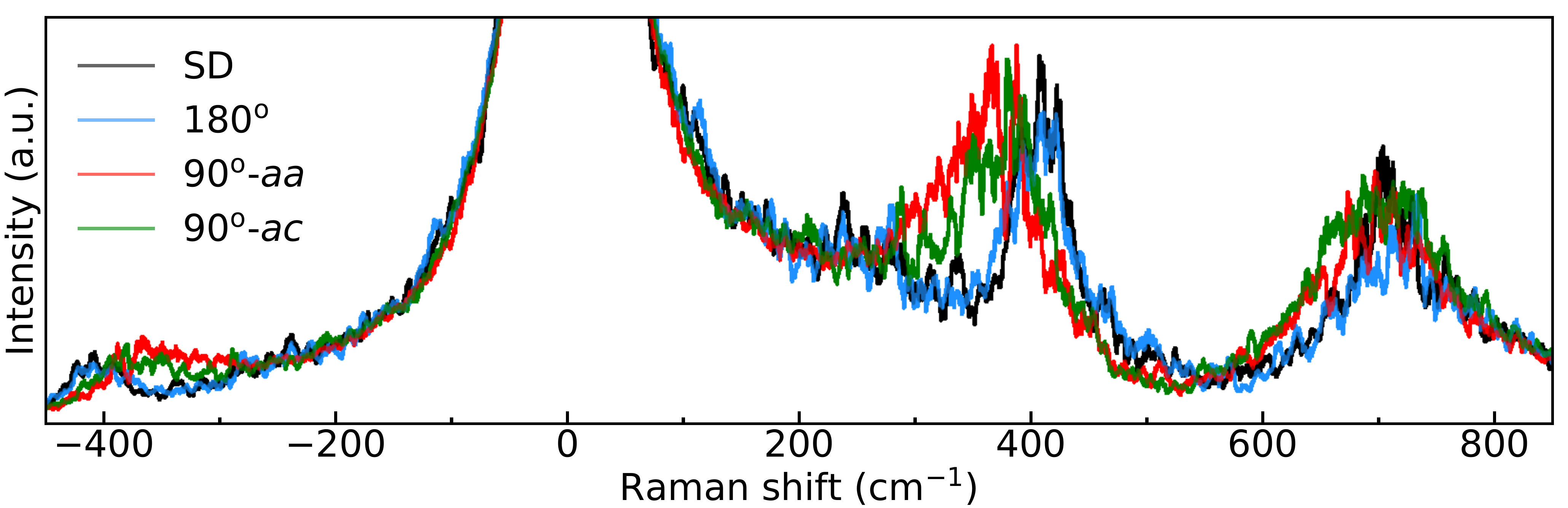}
\caption{(Color online) Raman spectra for different Domain structure of BTO. 
}
\label{Fig5}
\end{figure}

Next we apply the model to the classic ferroelectric BTO for which an atomistic potential is available~\cite{BTO_atomistic}, enabling large MD simulations. 
We use the $ab$ $initio$ MD $\pmb{\alpha}$ data as calculated using 10-atom BTO cell to fit the parameters in Eq.~\ref{eq1} for the TiO$_{6}$ octahedron only. We ignore the contribution of the BaO$_{12}$ polyhedra to $\pmb{\alpha}$ because it is small relative to that of TiO$_6$ due to predominantly Ti 3$d$ character of the conduction band  (see the SM for more information). In Fig.~\ref{Fig3} (a), we present the trajectories of $\alpha_{xx}$ at 300 K as calculated using DFT and the parameterized model, respectively. It is observed here as well that the model $\alpha_{xx}$ trajectory is in good agreement with the DFT trajectory. For direct comparison of the model with DFT results, we calculated Raman spectra using DFT $\alpha_{xx}$ and model $\alpha_{xx}$ with the results shown in Fig.~\ref{Fig3} (b) \footnote{Though this is not the actual spectra as we considered trajectory for small time with large time step, these results indicate  the accuracy of this model.}. 
The figure shows that the DFT-based and model-based spectra are in good  agreement. 

Next, we considered a large simulation cell of single-domain BTO with the dimensions of 16$\times$16$\times$16 (20,480 atoms)  in order to examine the accuracy of the model in characterizing the  phase transition of BTO.  We calculated the trajectories using atomistic MD with the bond-valence potential for different temperature between 10~K and 160~K which capture all of the phases of BTO for this potential \footnote{Since the BV potential is based on DFT calculations, it underestimates the Curie temperature of BTO due to the underestimation of the O$_6$ rotation energy cost by DFT as shown in previous work~\cite{BTO_atomistic}}. The evolution of polarization with time for all of these phases is presented in the SM. We then used the same constants as obtained for 10-atom BTO cell to calculate the Raman spectra of the single-domain 16$\times$16$\times$16 BTO for all the considered temperatures. Fig.~\ref{Fig4} compares the results obtained by our model calculations  with the experimentally obtained spectra of single-crystal BTO (see SM also). 
Examination of the theoretical Raman spectra shows that rhombohedral, orthorhombic, tetragonal and cubic phases are achieved at $<$80~K, 90-95~K, 100-130~K and $>$160~K, respectively. This corresponds to the experimentally observed phase transitions at 183, 278 and 393 K due to the underestimation of the phase transition temperatures by DFT-based potentials~\cite{BTO_atomistic}. Additionally, with the increase of temperature, both experimental and theoretical spectra show peak broadening.  For the rhombohedral phase at 40 K in the theoretical results, the peaks below 200 cm$^{-1}$ match well the R-phase  experimental peaks below 200 cm$^{-1}$. The two peaks in the range of 200-500 cm$^{-1}$ in the theoretical spectra correspond to the two peaks in the 200-400 cm$^{-1}$ in the experimental spectra. The peaks between 500 and 700 cm$^{-1}$ in  theoretical spectra correspond to the peak around 500 cm$^{-1}$ observed experimentally. The peak in the high wavenumber region as observed theoretically (see SM) is also consistent with the experimental peak above 700 cm$^{-1}$~\cite{C7TA11096K}. Similarly, the comparison between theoretical results and experimental results shows good agreement for other phases of BTO. We also applied the model to CPB halide perovskite and obtained good agreement between DFT and model polarizability trajectories and Raman spectra (see SM).

To further demonstrate the utility of the model, we applied it to the Raman spectra of different ferroelectric domain structures of BTO to show that Raman spectra can distinguish between  the different domain phases and these difference can be interpreted using the atomistic polarizability model.
We considered single-domain, 180$^{\circ}$ domain, 90$^{\circ}$($aa$) domain and 90$^{\circ}$($ac$) domain structures of BTO  based on MD simulations using 120$\times$120$\times$6 supercell (432,000 atoms) with the  obtained Raman spectra shown in Fig.~\ref{Fig5}. It is clear that different domain variants can be detected based on the differences between their respective Raman spectra. These simulations also show that Raman spectra can be obtained for large supercells at essentially negligible computational cost that is only $\sim$ 50\% of the computational cost of BV potential MD simulations. (70 CPU hours required for MD simulations and 35 CPU hours required for each Raman spectrum calculations on 32 core machine).

To summarize, we have demonstrated that a simple, physically-based atomistic model with only 10 adjustable parameters can accurately represent the changes in the electronic polarizability of complex system such as BTO. Due its physical basis, the model is much more compact than the polynomial expansion  (128 adjustable parameters for H$_2$O molecule~\cite{polynomial}) and the deep-learning models of the polarizability tensor.
This enables efficient calculations of Raman spectra from large-scale MD simulations with only modest computational resources as well as atomistic and local analysis of the contributions to Raman spectra for the interpretation of Raman spectroscopy results.

\bibliographystyle{apsrev4-1} 

\providecommand{\noopsort}[1]{}\providecommand{\singleletter}[1]{#1}%

\end{document}


\title{Supplemental Material: An atomistic model of electronic polarizability for calculation of Raman scattering from large-scale MD simulations}

\author{Atanu Paul$^1$}
\author{Anthony Ruffino$^3$}
\author{Stefan Masiuk$^2$}
\author{Jonathan Spanier$^{2,3}$}
\author{Ilya Grinberg$^1$}
\email[]{ilya.grinberg@biu.ac.il}

\affiliation{$^1$Department of Chemistry, Bar-Ilan University, Ramat Gan 5290002, Israel}
\affiliation{$^2$Department of Mechanical Engineering and Mechanics, Drexel University, Philadelphia, Pennsylvania 19104, USA}

\affiliation{$^3$Department of Physics, Drexel University, Philadelphia, Pennsylvania 19104, USA}

\maketitle
\vspace{3 cm}

\tableofcontents

\section{Computational details}
Density functional theory (DFT) as well as $ab$ $initio$ molecular dynamics (AIMD) calculations were performed using the Quantum ESPRESSO code~\cite{QE} with the Perdew-Burke-Ernzerhof exchange-correlation functional for
solids (PBEsol)~\cite{PBEsol} for all the systems considered here. The dielectric function 
was calculated within the framework of random phase approximation
(RPA) as implemented in Quantum ESPRESSO. The kinetic energy cut-off for the expansion of the wave functions was set to 40 and 110 Ry for the AIMD and dielectric function calculations, respectively, for all systems.

In order to calculate the Lorentzian constants for modeling of the Raman spectra of water clusters ((H$_{2}$O)$_{n}$: $n$ = 1-6, 8), we considered $n$ = 1, $i.e.$ H$_{2}$O molecule only. We then extracted eight constants (because the  asymmetry effect  is unnecessary for  H atoms) after optimizing the model polarizability with respect to that obtained from DFT calculations on AIMD structures (300 structures) at 300 K. The extracted values of the Lorentzian constants are $a_{1}$ = 0.163408, $a_{2}$ = -0.176637,
$a_{3}$ = 0.640497, $a_{4}$ = -0.325206, $a_{5}$ = -2.420756, $a_{6}$ = 0.667671, $a_{7}$ = 2.626936, $a_{8}$ = 6.445961. Those constants were used to calculate the atomistic model polarizability for other (H$_{2}$O)$_{n}$ systems. 
Classical molecular dynamics of (H$_{2}$O)$_{n}$ was performed using MB-pol~\cite{MBpol1,MBpol2} to obtain a long trajectory of up to 3 ns with a time step of 0.001 ps, where the temperature of the dynamics were 50 K for $n$ = 2 to 6, and 10 K for $n$ = 1  and 8. In order to compare the model spectra with DFT spectra for H$_{2}$O, (H$_{2}$O)$_{2}$ and (H$_{2}$O)$_{3}$, a short trajectory of up to 44 ps with a time step of 0.003 ps was also considered.

The Lorentzian constants for BaTiO$_{3}$ (BTO) were extracted after optimizing the model polarizability with respect to polarizability from DFT where the structures for the DFT calculations were taken from AIMD on 10-atom BTO cell  at 300 K, 600 K, 800 K and 1000 K. After optimizing
the parameters to reproduce the polarizabilities of 120 structures (30 from each temperature), the extracted values were obtained as $a_{1}$ = 1.543787, $a_{2}$ = 2.030938,
$a_{3}$ = 1.635022, $a_{4}$ = 3.113967, $a_{5}$ = -0.481618, $a_{6}$ = 3.091547, $a_{7}$ = -1.060428, $a_{8}$ = 6.501974, $a_{9}$ = 0.227096 and $a_{10}$ = 2.492371. While performing the AIMD, the Brillouin zone integration was executed using 2$\times$4$\times$4 and 4$\times$4$\times$4 \textbf{\emph{k}} mesh for 2$\times$1$\times$1 (10 atoms) and 1$\times$1$\times$1 (5 atoms) cells of BTO, respectively. To calculate the polarizability tensor from DFT, we considered a larger \textbf{\emph{k}} mesh of 3$\times$6$\times$6 and 6$\times$6$\times$6 for the 2$\times$1$\times$1 (10 atoms) and 1$\times$1$\times$1 (5 atoms) cells of BTO, respectively. The classical MD simulations were performed using the LAMMPS code~\cite{LAMMPS} with the atomistic potential~\cite{BTO_atomistic} for BTO  for time up to 300 ps with a time step 0.001 ps to obtain dynamics of BTO single-domain and polydomain systems. To simulate single-crystal single domain BTO, we considered a  16$\times$16$\times$16 cell size. The dynamics of the BTO domain structures (single (SD), 180$^{0}$, 90$^{0}$-$aa$ and 90$^{0}$-$ac$ domain) were investigated using the  120$\times$120$\times$6 cells.

The constants for the Lorentzian functions in the atomistic polarizability model for  CsPbBr$_{3}$ (CPB) were evaluated in a similar manner to those for BTO and were derived from AIMD and DFT results obtained for the 2$\times$1$\times$1 (10 atoms) cell of CBP at 300 K, 600 K and 800 K. The values of the Lorentzian constants for CPB are $a_{1}$ = -2.433495, $a_{2}$ = 0.799679,
$a_{3}$ = 3.013168, $a_{4}$ = 1.636966, $a_{5}$ = 6.067340, $a_{6}$ = 2.496868, $a_{7}$ = -13.071122, $a_{8}$ =  9.703365, $a_{9}$ = 0.615866, $a_{10}$ = 6.620343. The Brillouin integration was performed using the similar size of \textbf{\emph{k}} mesh as already described for BTO. In order to compare the Raman spectra from model and DFT calculation, we considered a MD trajectory of short time of 60 ps with a large time step of 0.15 ps.  

To extract Lorentzian constants for CO$_{2}$, the polarizability model parameters were optimized to match the polarizability calculated by DFT for 200 structures obtained from AIMD at 300 K. The extracted Lorentzian constants for CO$_{2}$ molecule are $a_{1}$ = 0.311722, $a_{2}$ = 1.067479, $a_{3}$ = 1.064764, $a_{4}$ = 0.255176, $a_{5}$ = 0.160768, $a_{6}$ = 1.316444, $a_{7}$ = 0.940256 and $a_{8}$ = 1.158281. However, the Raman spectrum of CO$_{2}$ was calculated for a trajectory of 64 ps with a time step of 0.0014 ps as obtained from AIMD at 300 K.

It is important to note that all the extracted Lorentzian constants are not unique  and may vary depending on the optimization conditions.
\begin{figure*}[]
\centering
\includegraphics[width = 16 cm, angle =0 ]{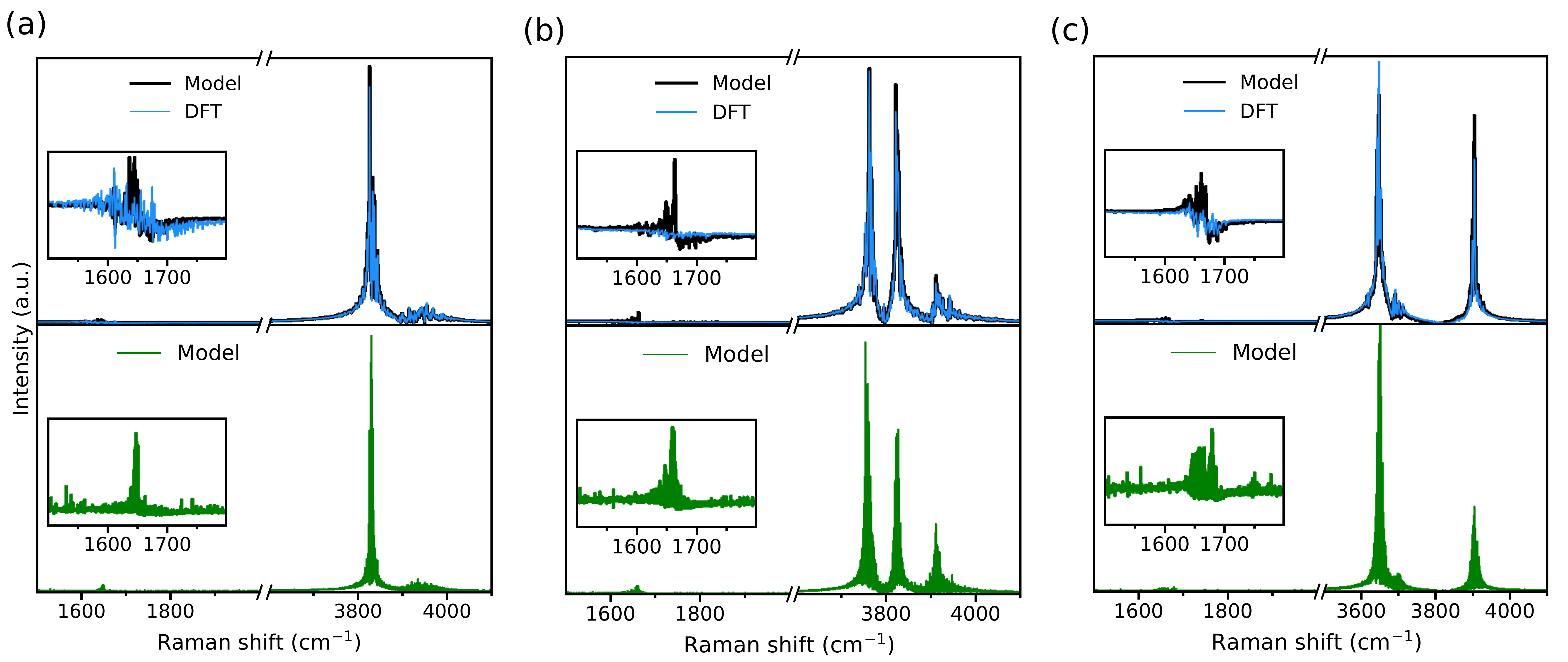}
\caption{(color online) Raman spectra of (a) H$_{2}$O: for short
trajectory (top panel) and for long trajectory (bottom panel), (b) (H$_{2}$O)$_{2}$: for short trajectory (top panel) and for long trajectory (bottom panel), (c) (H$_{2}$O)$_{3}$: for short trajectory (top panel) and for long trajectory (bottom panel). Model spectra in blue and green. DFT spectra in filled red. Inset of each graph shows the spectra in the region of 1600 cm$^{-1}$.}
\label{H2O_Raman_123} 
\end{figure*}
\section{Experimental details}
Inelastic light scattering (Raman scattering) was performed in the backscattering VV configuration ($Z(X,X)\bar{Z}$, X$\parallel$(100)) with a 5 mW 385 nm excitation source (M Squared Lasers) focused to a surface intensity of 1 W/cm$^2$. Light was dispersed by a T64000 triple spectrometer (Horiba Jobin-Yvon), and collected with an air-cooled CCD (Andor). The (001)-oriented BaTiO$_3$ bulk single crystal was grown by top seeded solution growth (MTI Corp.).

Raman spectra were collected between 123 K and 423 K at 5 K intervals, and each individual spectrum is 3 averaged acquisitions of 10 seconds. Experimental observations of abrupt structural phase transitions at 183 K (R to O), 278 K (O to T), and 393 K (T to C) are in agreement with existing literature on the material. Additionally, the intensity of the collapsing soft mode (160 cm$^{-1}$ in R-phase, and the broad feature at 35 cm$^{-1}$ in O-phase) are an indication of structural homogeneity and a low population of domain walls, showing these are effectively single-domain measurements.

\section{H$_{2}$O clusters}
The Raman spectra from model and DFT for the cases of $n$ = 1, 2 and 3 using short trajectory are already presented in the main paper. In Fig. S\ref{H2O_Raman_123} (a), (b), (c), we plot the Raman spectra again but using long trajectory of 3 ns and smaller time step of 0.001 ps along with the spectra obtained using the short trajectory for $n$ = 1, 2 and 3. The long trajectory clearly also find the similar peak position as obtained using short trajectory.  

\begin{figure}[]
\centering
\includegraphics[width = 8.3 cm, angle =0 ]{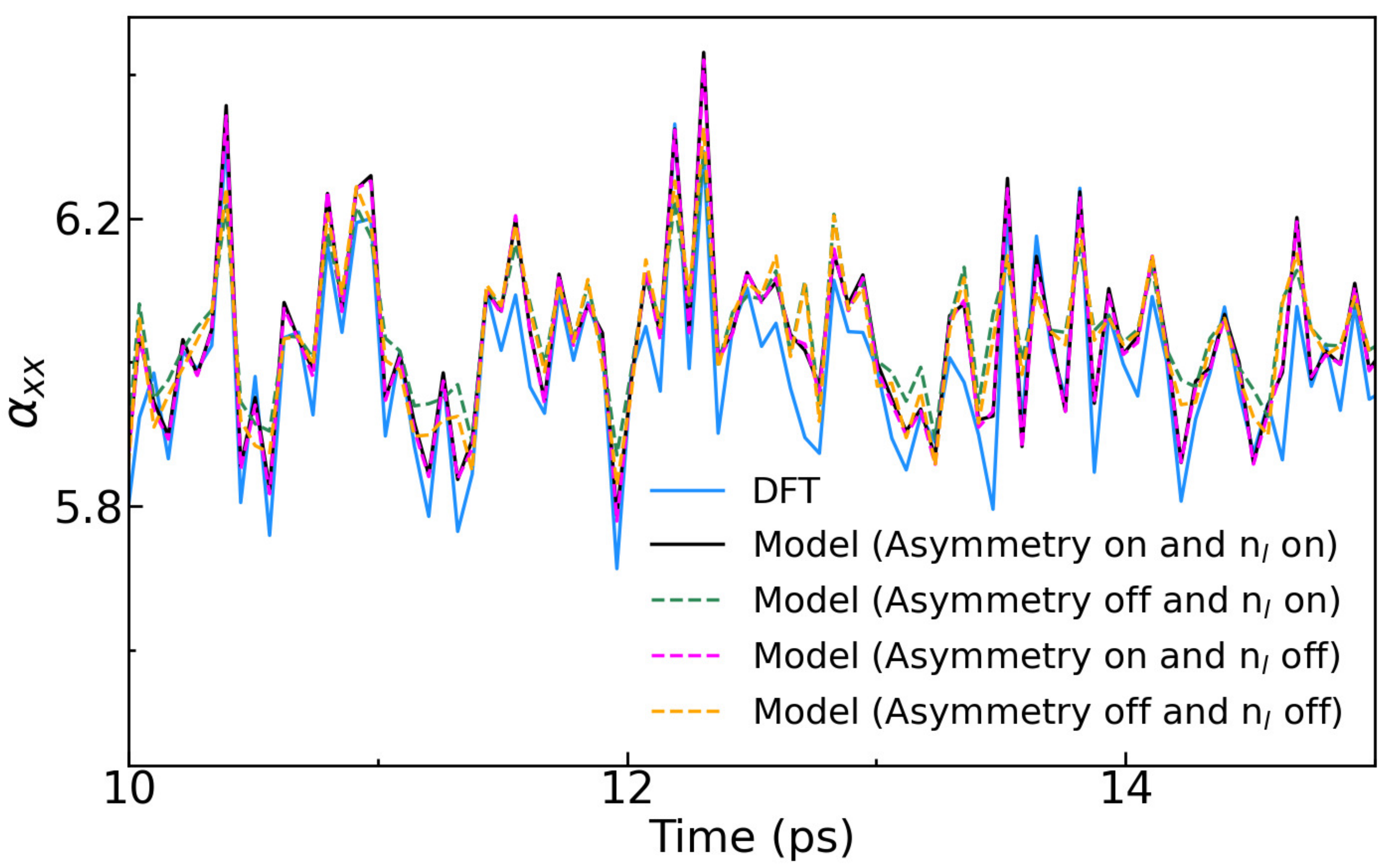}
 \caption{(color online) Comparison of the polarizability ($\alpha_{xx}$) trajectories from DFT, model (asymmetry on and $m$ on), model (asymmetry off  and $m$ on), model (asymmetry on and $m$ off) and model (asymmetry off and $m$ off) for BTO (5 atom cell) at 300 K.}
\label{n_n_asym_check} 
\end{figure}
\begin{figure}[]
\centering
\includegraphics[width = 8.4 cm, angle =0 ]{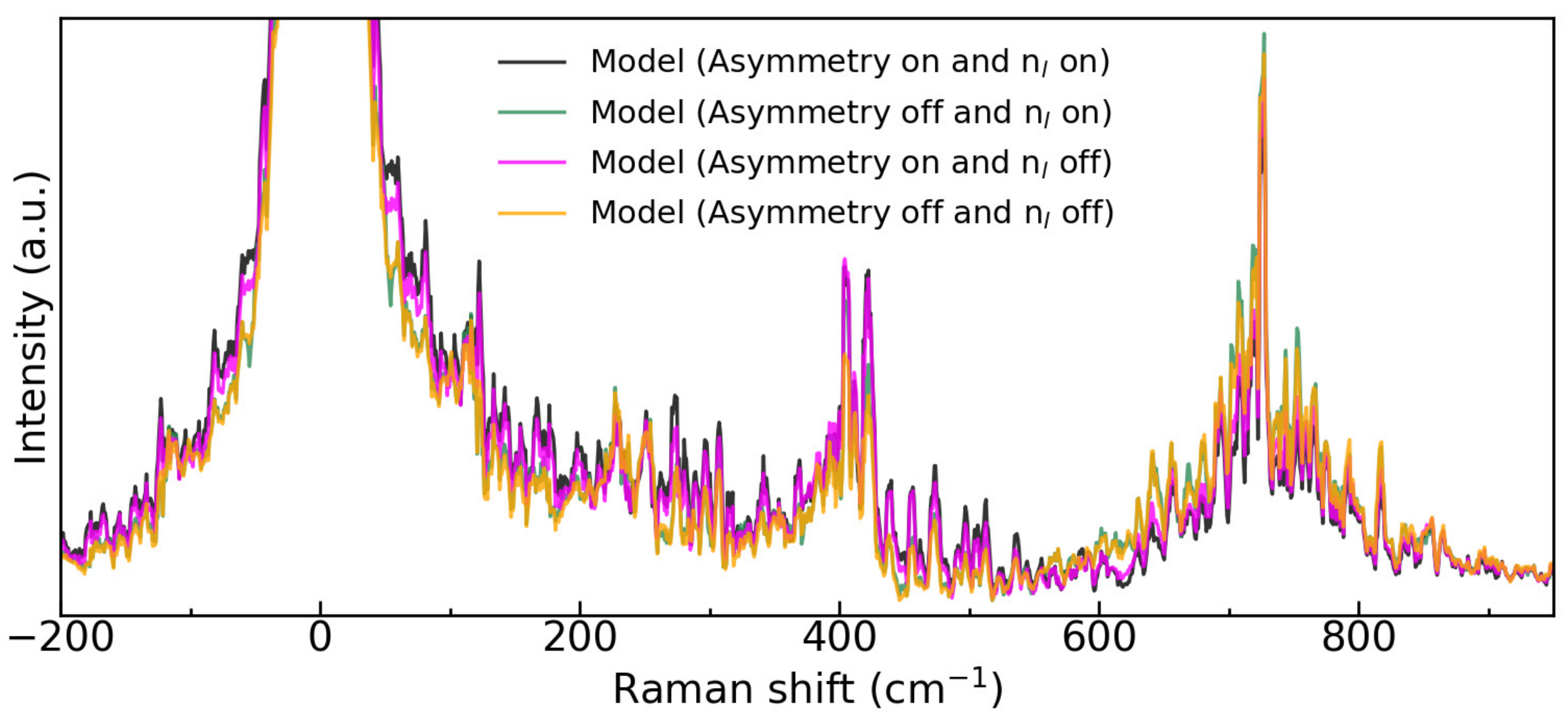}
 \caption{(color online) Comparison of Raman spectra for single domain of BTO thin film (cell size: 30$\times$10$\times$10) using the polarizability trajectory from the model (asymmetry on and $m$ on), model (asymmetry off  and $m$ on), model (asymmetry on and $m$ off) and model (asymmetry off and $m$ off).}
\label{Raman_301010} 
\end{figure}
\begin{figure}[]
\centering
\includegraphics[width = 8 cm, angle =0 ]{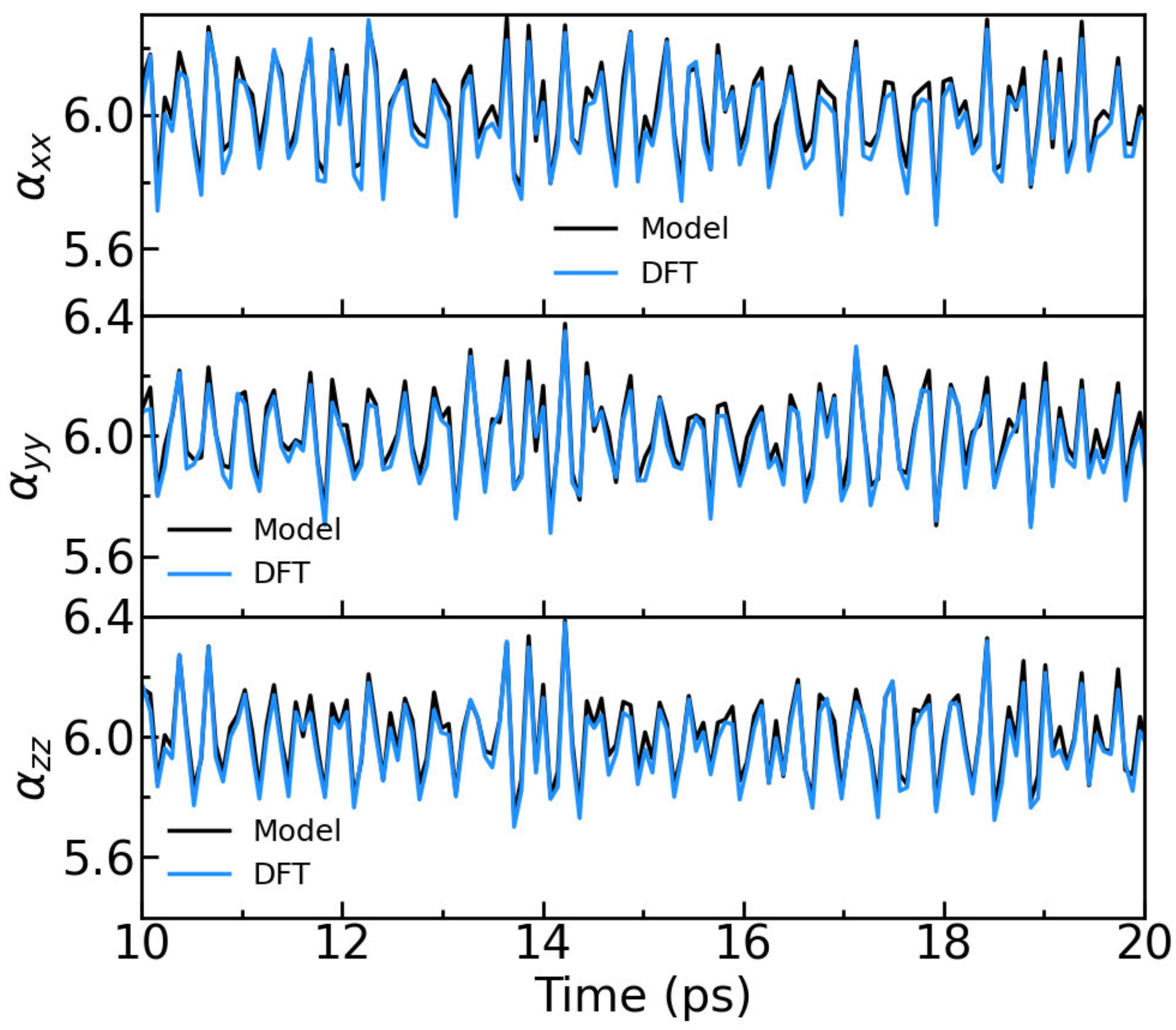}
 \caption{(color online) Polarizability ($\alpha_{xx}$ (top panel), $\alpha_{yy}$ (middle panel) and $\alpha_{zz}$ (bottom panel)) trajectory from DFT and model of BTO at 200 K.}
\label{BTO_200_5} 
\end{figure}
\begin{figure}[]
\centering
\includegraphics[width = 8 cm, angle =0 ]{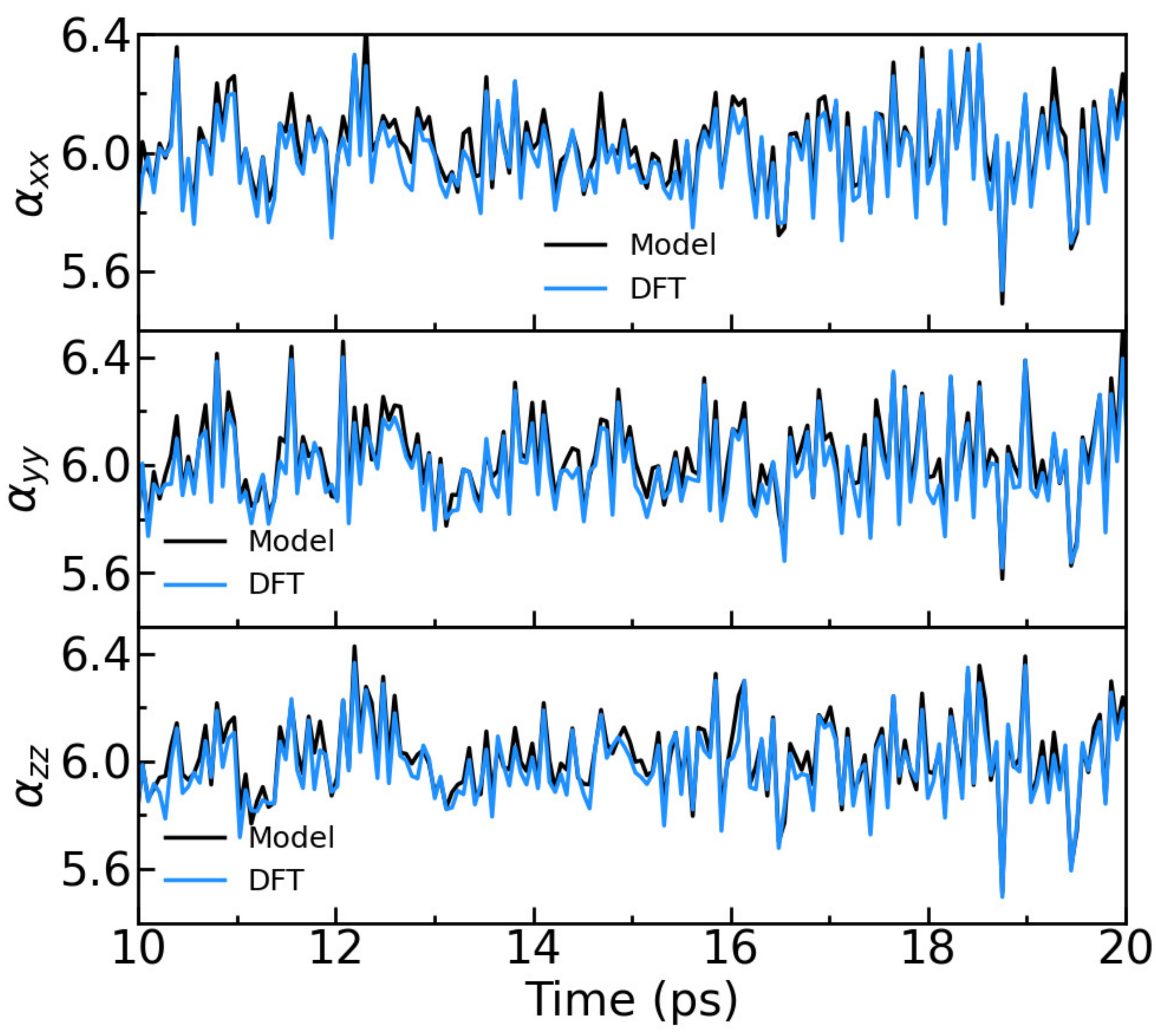}
 \caption{(color online) Polarizability ($\alpha_{xx}$ (top panel), $\alpha_{yy}$ (middle panel) and $\alpha_{zz}$ (bottom panel)) trajectory from DFT and model of BTO at 300 K.}
\label{BTO_300_5} 
\end{figure}
\begin{figure}[]
\centering
\includegraphics[width = 8 cm, angle =0 ]{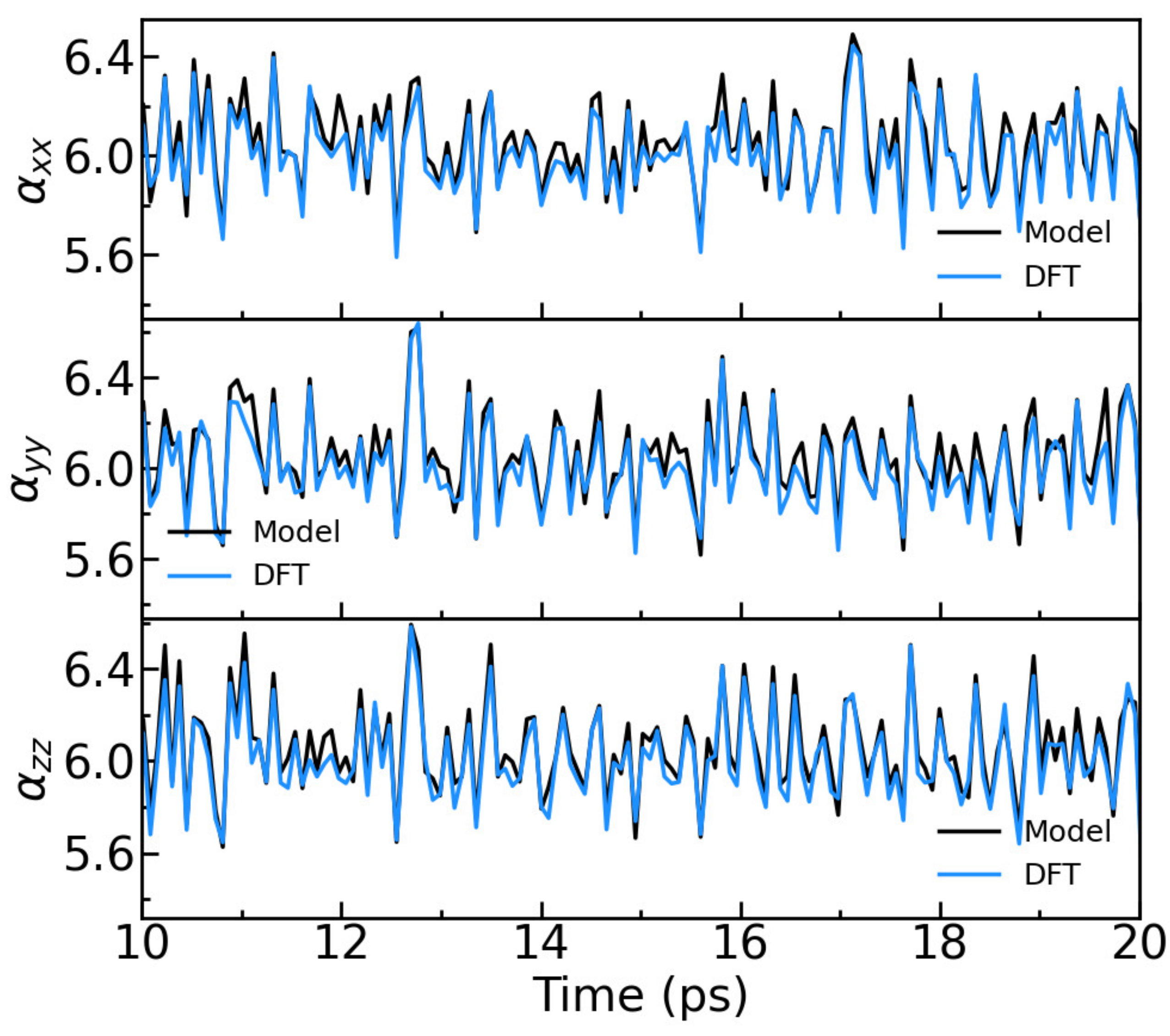}
 \caption{(color online) Polarizability ($\alpha_{xx}$ (top panel), $\alpha_{yy}$ (middle panel) and $\alpha_{zz}$ (bottom panel)) trajectory from DFT and model of BTO at 400 K.}
\label{BTO_400_5} 
\end{figure}
\begin{figure}[]
\centering
\includegraphics[width = 8 cm, angle =0 ]{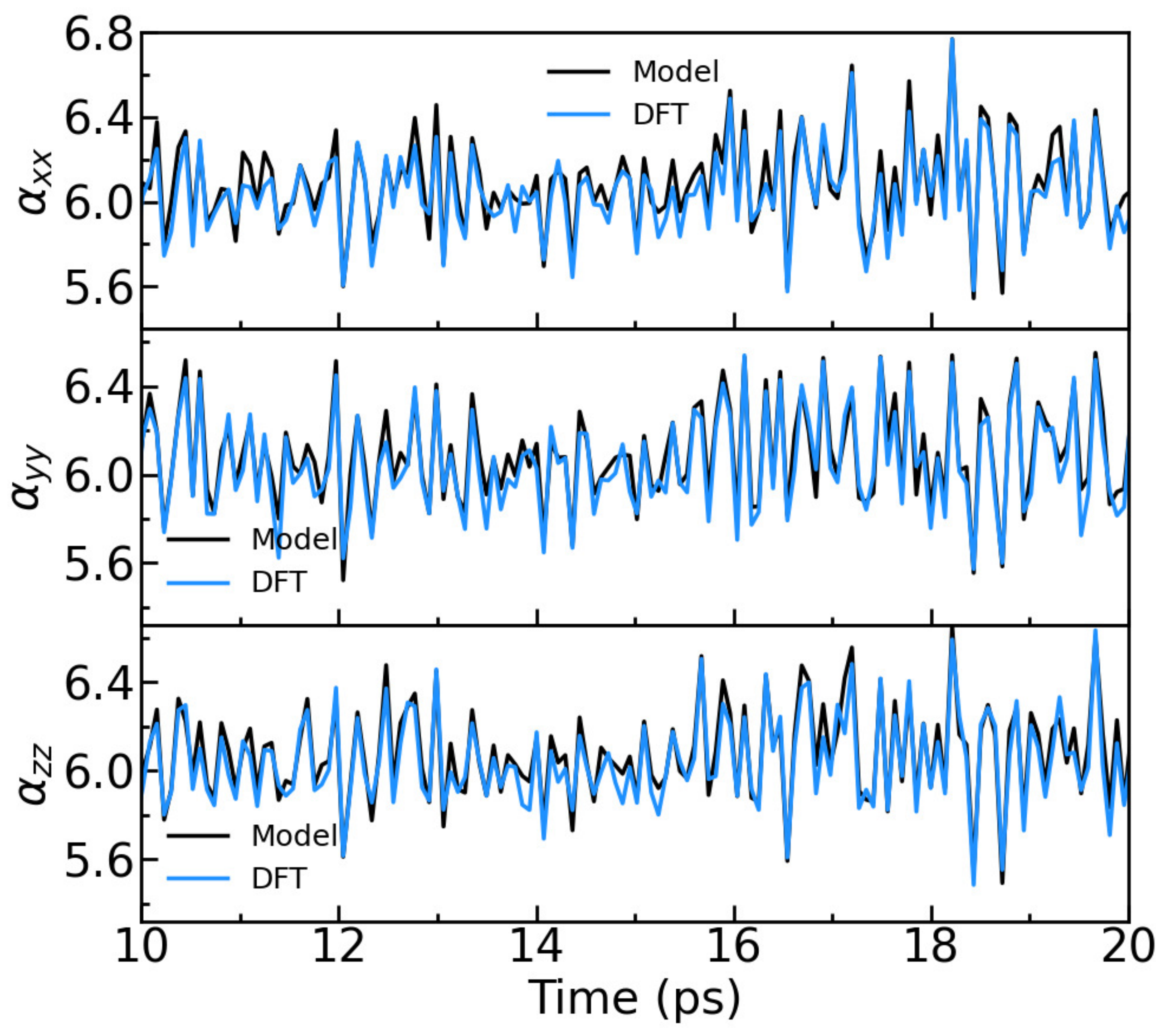}
 \caption{(color online) Polarizability ($\alpha_{xx}$ (top panel), $\alpha_{yy}$ (middle panel) and $\alpha_{zz}$ (bottom panel)) trajectory from DFT and model of BTO at 600 K.}
\label{BTO_600_5} 
\end{figure}
\begin{figure}[]
\centering
\includegraphics[width = 8 cm, angle =0 ]{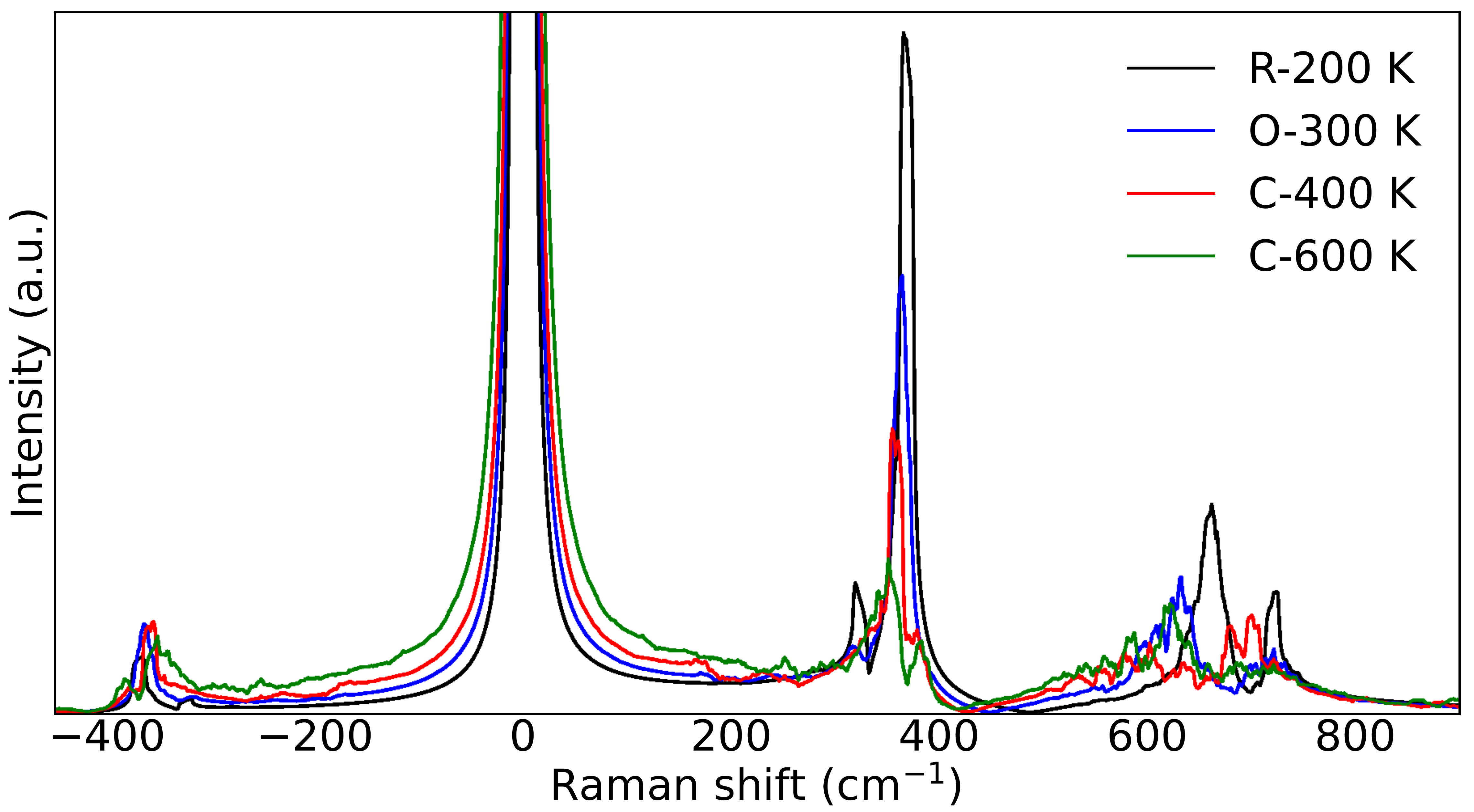}
 \caption{(color online) Raman spectra of 5-atom BTO for different temperatures. The temperature along with the phases (R: rhombohedral, O: orthorhombic, C: cubic) are mentioned.}
\label{spectra_5} 
\end{figure}
\begin{figure*}[h!]
\centering
\includegraphics[width = 17.8 cm, angle =0 ]{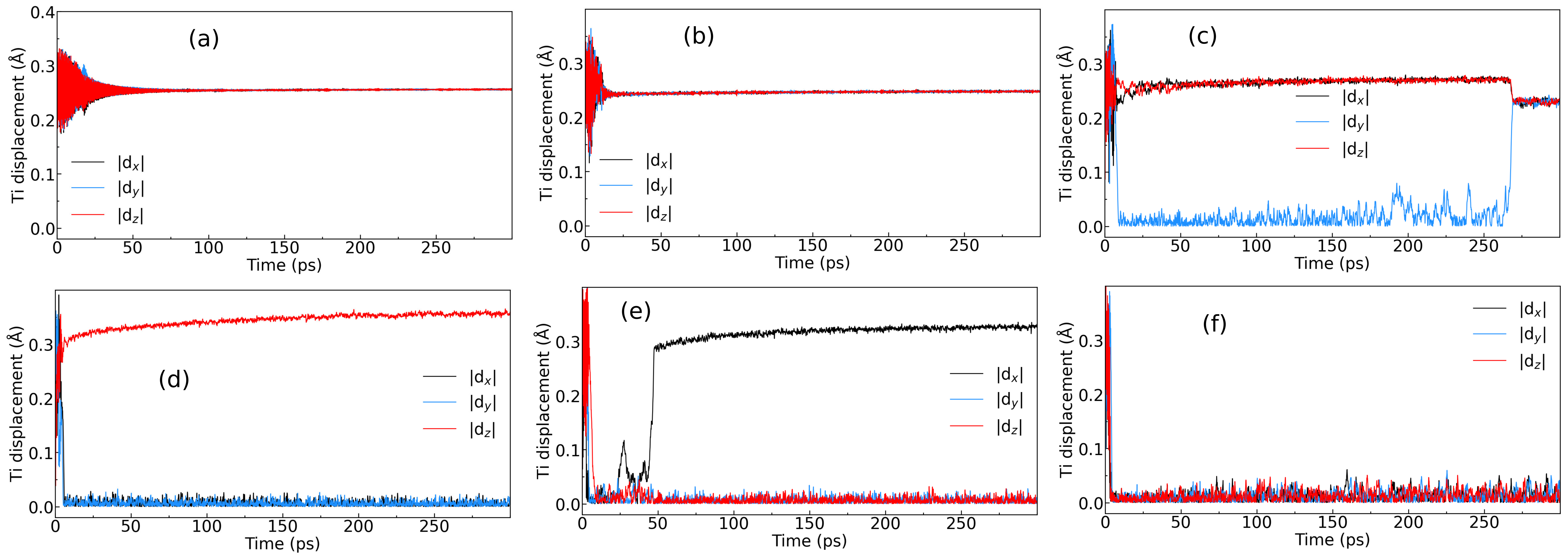}
 \caption{(color online) Ti displacement (\textbf{d}) for up to 300 ps at temperatures of (a) 20 K (rhombohedral phase), (b) 50 K (rhombohedral phase), (c) 92 K (orthorhombic phase: up to 260 ps, rhombohedral phase: beyond 270 ps), (d) 110 K (tetragonal phase), (e) 130 K (tetragonal phase: above 50 ps) and (f) 160 K (cubic phase).}
\label{Ti_dis_161616} 
\end{figure*}
\begin{figure*}[h!]
\centering
\includegraphics[width = 17.8 cm, angle =0 ]{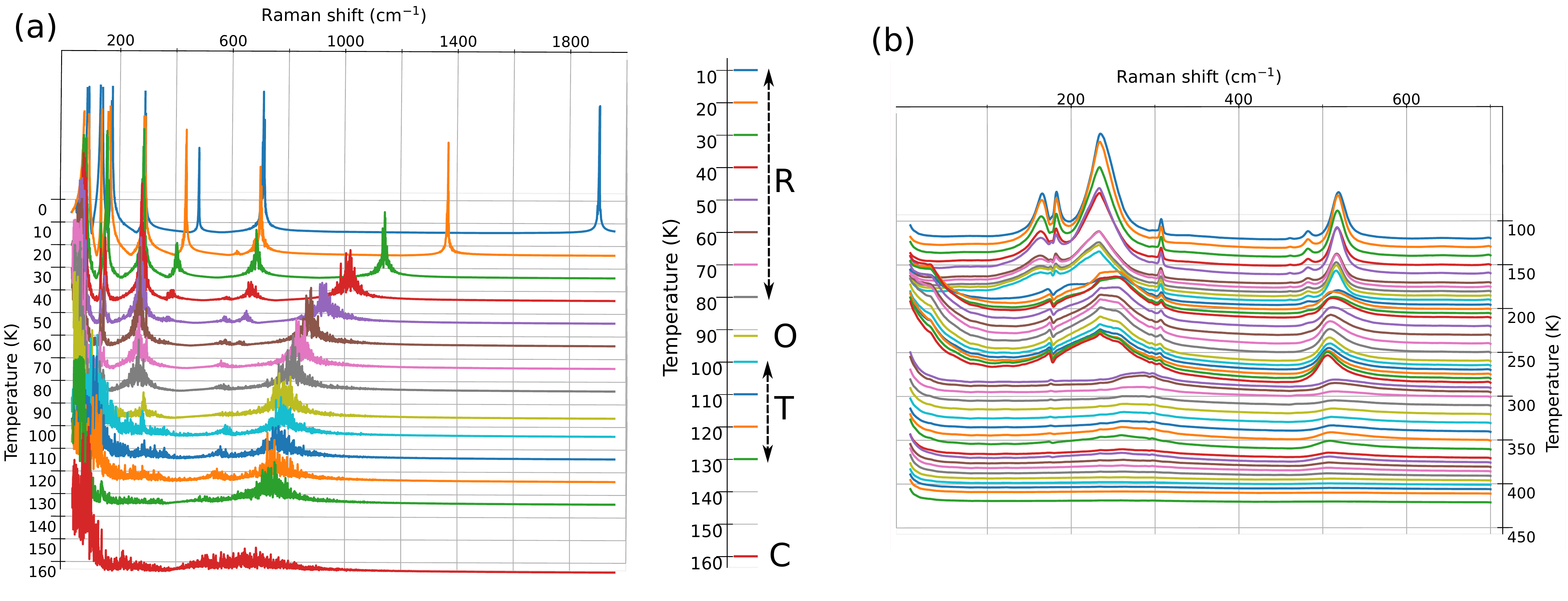}
\caption{(color online) Raman spectra of single-domain BTO for different temperatures as shown from (a) model and (b) experiment. R, O, T and C in the color bar represent the rhombohedral, orthorhombic, tetragonal and cubic phase for model.}
\label{161616_Raman} 
\end{figure*}
\begin{figure*}[h!]
\centering
\includegraphics[width = 17 cm, angle =0 ]{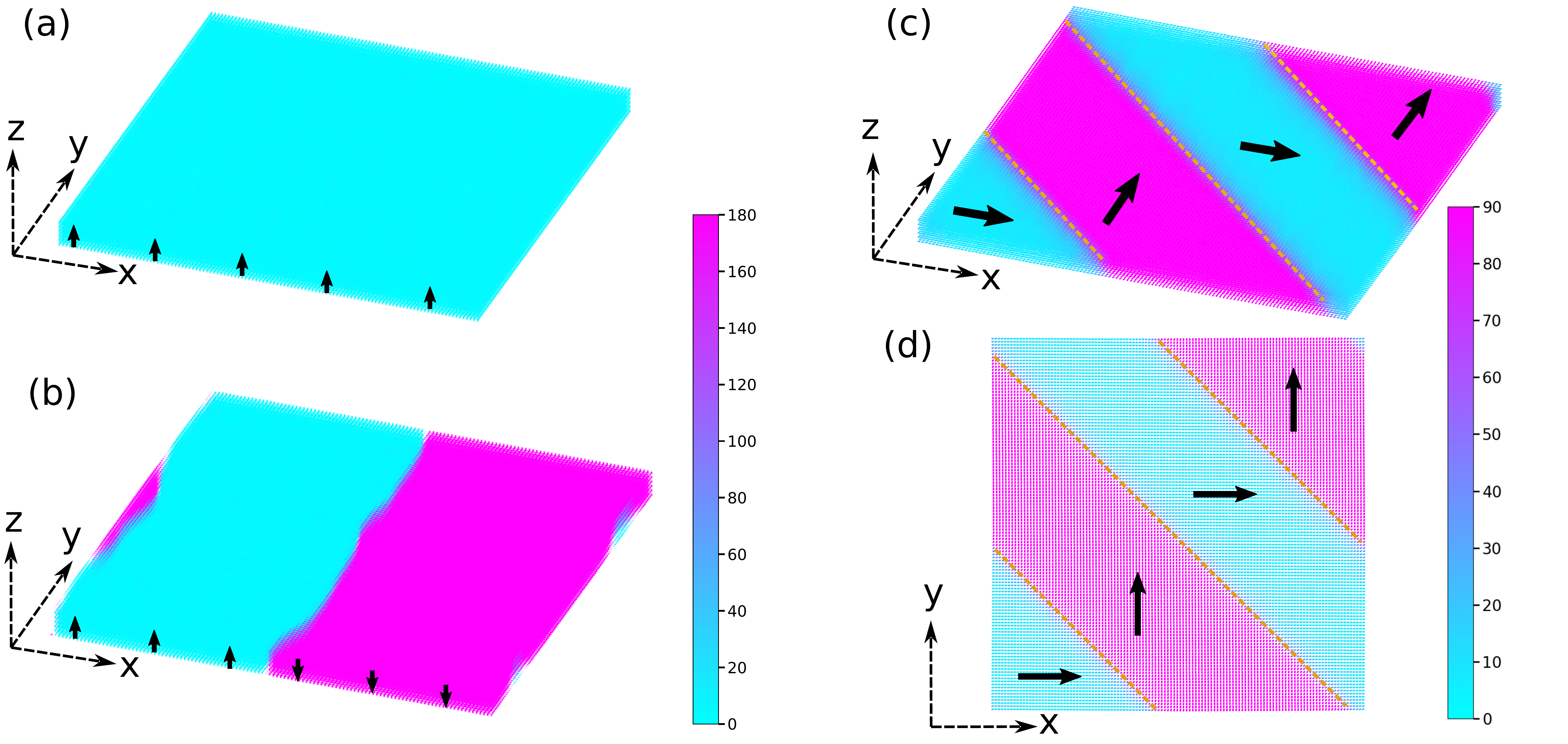}
 \caption{(color online) (a) single domain, (b) 180$^{0}$ domain, (c) $ac$ or $aa$ 90$^{0}$ domain, (d) 2D projection of $ac$ or $aa$ 90$^{0}$ domain structure of BTO (cell size: 120$\times$120$\times$6). The arrows (black) represent the directions of polarization in each domain. The orange dotted lines indicate the domain wall region in case of (c) and (d).}
\label{domain} 
\end{figure*}
\section{Importance of asymmetry and dynamical charge state ($m$)}
The expression for the polarizability (Eq. 2 of the main paper) is composed of symmetric and asymmetric parts where the symmetric part also contain the dynamical charge state ($m$). 
In this section, the importance of each term for the calculation of the spectra is examined. We considered BTO and evaluate the polarizability as well as Raman spectra with the model for the following four cases: (a) the effects of both asymmetry and $m$ are included, (b) the effect of asymmetry is not included but the effect of $m$ is included, (c) the effect of asymmetry is included but the effect of $m$ is not included, (d) the effects of both asymmetry and $m$ are not included. The Lorentzian constants for case (a) are already presented in the \textquoteleft COMPUTATIONAL DETAILS\textquoteright\  section. The extracted Lorentzian constants for the other cases are (b) $a_{1}$ = -3.411483, $a_{2}$ = -1.503017,
$a_{3}$ = 1.392625, $a_{4}$ = 1.976437, $a_{5}$ = -1.092110, $a_{6}$ = 0.932035, (c) $a_{1}$ = 1.228825, $a_{2}$ = 2.203616, $a_{3}$ = 1.089328, $a_{4}$ = 2.380694, $a_{5}$ = 0.003688, $a_{6}$ = 2.304688, $a_{7}$ = -2.924001, $a_{8}$ = 3.447771, $a_{9}$ = 0.243666, $a_{10}$ = 2.528604, (d) $a_{1}$ = -2.236458, $a_{2}$ = -0.183282,
$a_{3}$ = 1.235218, $a_{4}$ = 1.469221, $a_{5}$ = -0.336989, $a_{6}$ = 0.944660. Fig. S\ref{n_n_asym_check} represents the polarizability trajectory calculated from the model for the four cases along with the DFT trajectory  for five-atom BTO at 300 K. For all cases, the model trajectory captured the DFT trajectory well indicating that the symmetric terms are the most important in the expression of model polarizability (Eq. 2 of the main paper). The trajectory of the models without the asymmetry terms show some deviation from the DFT trajectory, indicating that  the asymmetry terms are the next most important after the symmetric terms. By contrast, a careful observation of the figure shows that the inclusion  of $m$ in the model has a weak effect because the trajectories calculated from conditions (a) and (c) are nearly identical.  However, the importance of $m$ has already been demonstrated in the main paper for large lattice parameter changes.

We then used those four sets of constants as extracted for the above four cases in order to calculate the Raman spectra of single domain of BTO thin film (cell size: 30$\times$10$\times$10) at 130 K. Fig. S\ref{Raman_301010} shows the calculated Raman spectra for these four cases. For all cases,   similar peak positions are obtained in the spectra. The impact of each terms on the intensity are also visible in the spectra.

\section{BaTiO$_{3}$ five atoms cell}
Figs. S\ref{BTO_200_5}, S\ref{BTO_300_5}, S\ref{BTO_400_5} and S\ref{BTO_600_5} show the model polarizability ($\alpha_{xx}$, $\alpha_{yy}$ and $\alpha_{zz}$) trajectory along with
that from DFT in case of five-atom BTO at 200 K, 300 K, 400 K and 600 K, respectively, as obtained from AIMD, DFT and model. All the model trajectories show good agreement with the DFT results.

The Raman spectra at each temperature are also shown in Fig. S\ref{spectra_5}. The Raman spectra were calculated using AIMD trajectory of up to 25 ps with a time step of 0.0014 ps. 

\section{BaTiO$_{3}$ single domain and polydomain state calculations}

In Fig. S\ref{Ti_dis_161616}, we present the BTO phases by showing the time evolution of Ti displacement for up to 300 ps at each temperature after molecular dynamics simulation on single-domain BTO with a 16$\times$16$\times$16 cell. Figs. S\ref{Ti_dis_161616}(a) and (b) are obtained from the simulations at 20 K and 50 K, respectively, which correspond to the rhombohedral phase of BTO. Fig. S\ref{Ti_dis_161616}(c) is plotted for simulation temperature 92 K which shows the orthorhombic phase below 260 ps and rhombohedral phase above 270 ps. The molecular dynamics simulations at 110 K and 130 K captured the tetragonal phase as shown in Fig. S\ref{Ti_dis_161616}(d) and (e), respectively. The cubic phase is obtained at 160 K as shown in Fig. S\ref{Ti_dis_161616}(f). The Raman spectra of all of these phases have already been shown in the main paper. For the sake of completeness again we have shown the model Raman spectra along with the experimental Raman spectra in more details for the entire frequency range (see Fig. S\ref{161616_Raman}).\\    

In Fig. S\ref{domain}, we represents the various domain structures of BTO that can be achieved in thin films as obtained from the classical molecular dynamics using a 120$\times$120$\times$6 cell of BTO at 130 K. Figs. S\ref{domain} (a), (b) and (c) show the single-domain, and 180$^{0}$ and $aa$ or $ac$ 90$^{0}$ domain structures of BTO, respectively. The 2D projection of the $aa$ or $ac$ domain wall structure is also shown in Fig. S\ref{domain} (d).
\begin{figure}[]
\centering
\includegraphics[width = 8.5 cm,angle =0]{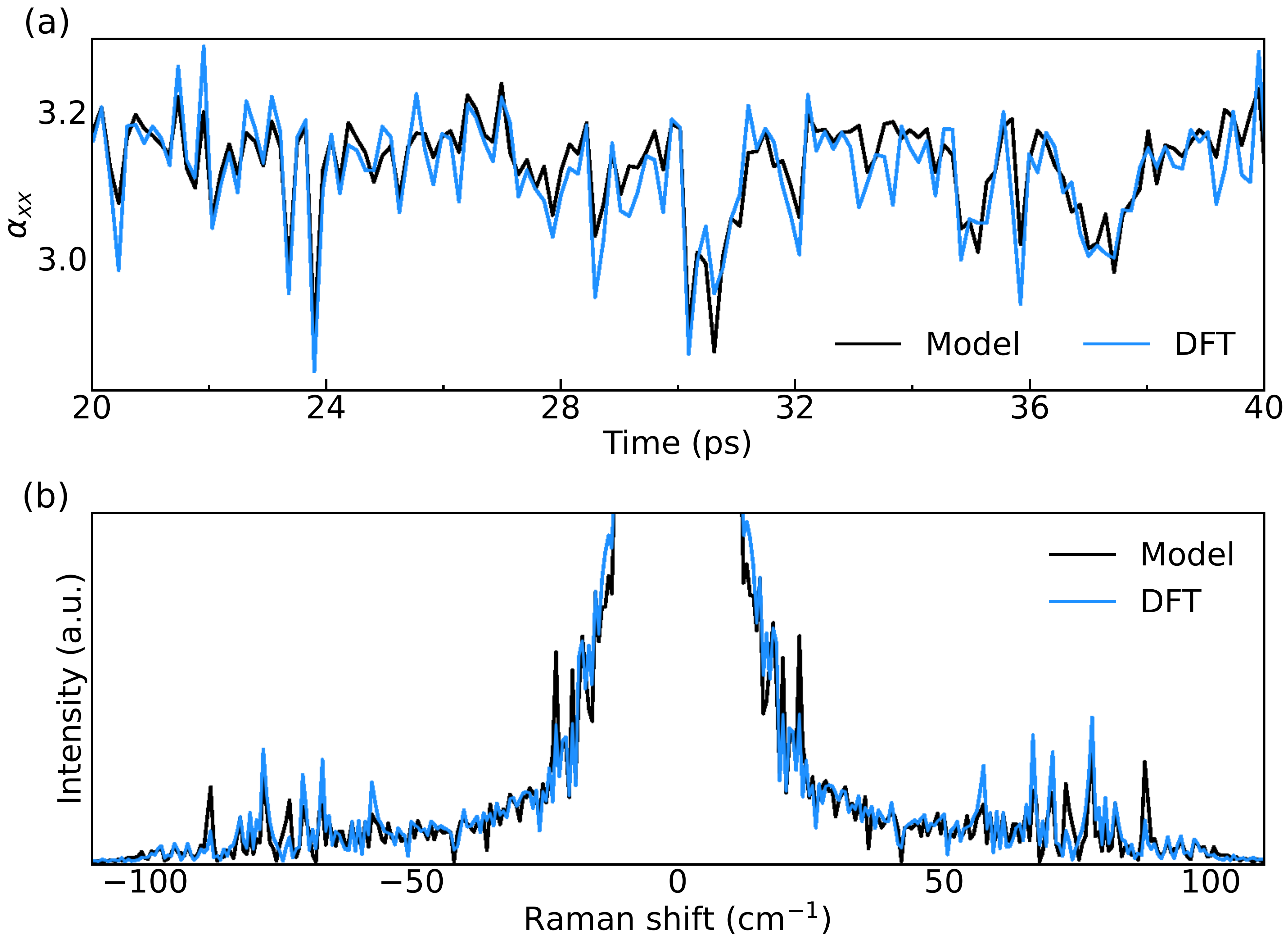}
\caption{(Color online) (a) Polarizability trajectory of CsPbBr$_{3}$ at 400 K using DFT and model. (b) Raman spectra CsPbBr$_{3}$ using the trajectory at 400 K.}
\label{Fig6}
\end{figure}

\begin{figure}[]
\centering
\includegraphics[width = 8.3 cm, angle =0 ]{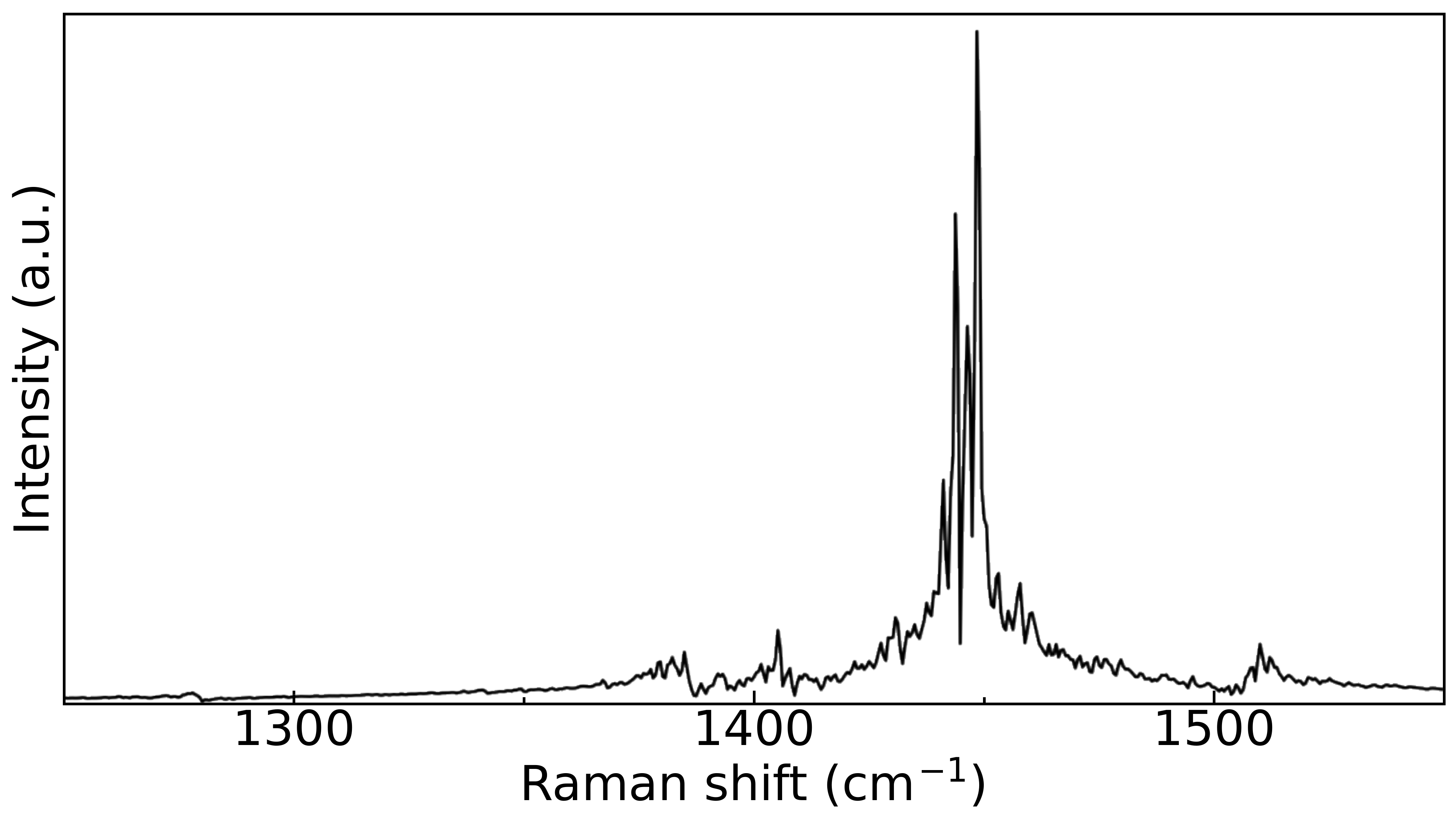}
 \caption{(color online) Raman spectra of CO$_{2}$ at 300 K using the model polarizability trajectory.}
\label{CO2_Raman} 
\end{figure}
\begin{figure}[]
\centering
\includegraphics[width = 8 cm, angle =0 ]{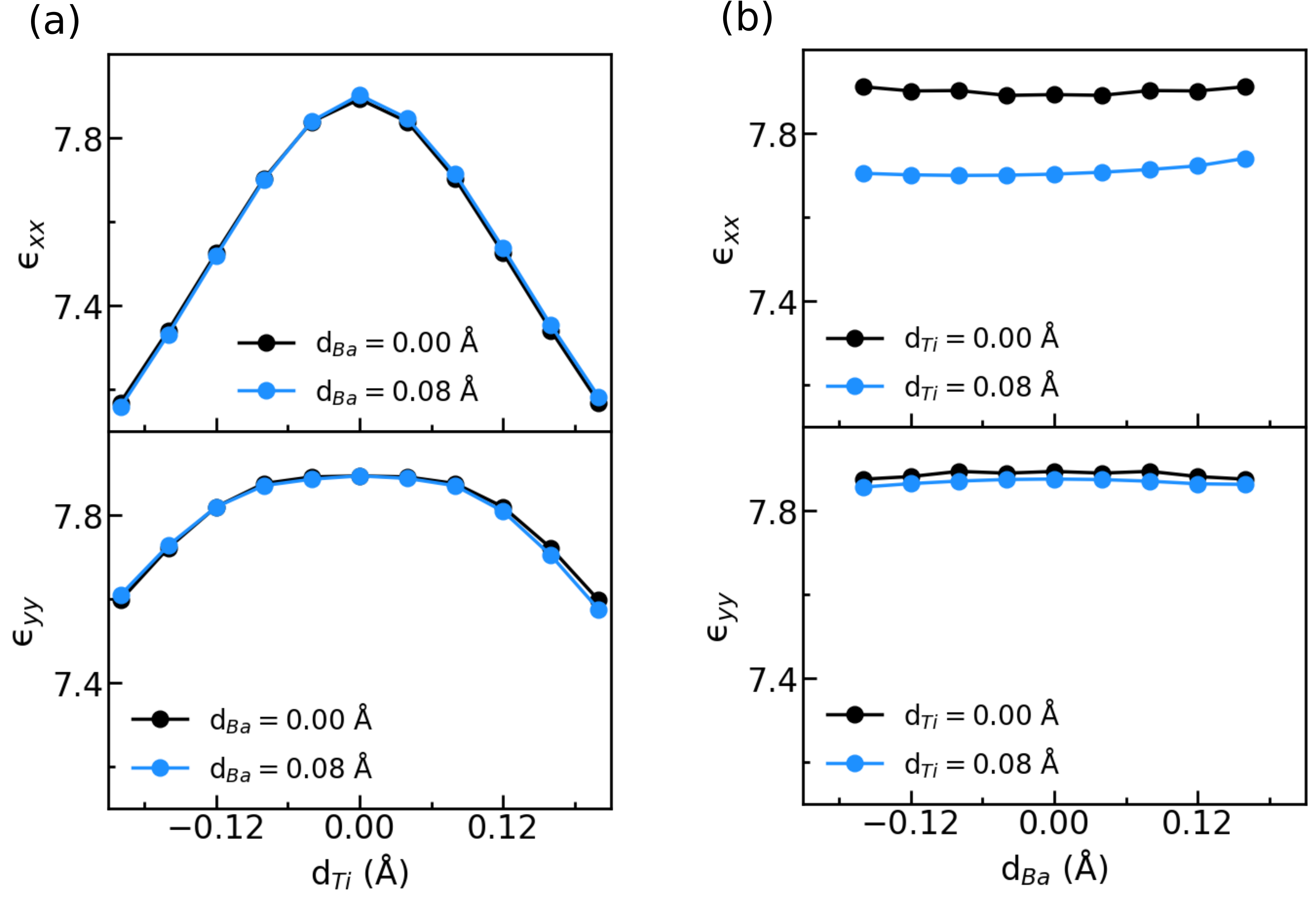}
 \caption{(color online) $\epsilon_{xx}$ and $\epsilon_{yy}$ of BTO (5-atom cell), where $d_{Ba}$ and $d_{Ti}$ are the displacements of Ba and Ti atoms respectively along $x$ direction from its centrosymmetric position denoted by $d_{Ba}$ = 0 and $d_{Ti}$ = 0. (a) $\epsilon_{xx}$ (first panel) and $\epsilon_{yy}$ (second panel) as a function of $d_{Ti}$ for two values of $d_{Ba}$ (0.00 and 0.08 \AA). (b) $\epsilon_{xx}$ (first panel) and $\epsilon_{yy}$ (second panel) as a function of $d_{Ba}$ for two values of $d_{Ti}$ (0.00 and 0.08 \AA).}
\label{Ba_atom_neglect} 
\end{figure}
\section{CsPbBr$_{3}$}
To demonstrate that the model is general and applicable  to non-oxide systems, we apply it to the halide perovskite CPB. The values of the constants are presented in 'COMPUTATIONAL DETAILS' section. In Fig. S\ref{Fig6} (a) we present the polarizability trajectory as calculated from model along with that from DFT for CPB at 400 K. Clearly, the model results reproduces the DFT results very well. Next,  we compared the model spectra with the DFT-derived spectra for short trajectory and large time step as shown in Fig. S\ref{Fig6} (b), confirming that this model can reproduce the  Raman spectra of CPB.
\section{CO$_{2}$ molecule}
The model Raman spectra of CO$_{2}$ at 300 K as calculated using AIMD trajectory for 64 ps is shown in Fig. S\ref{CO2_Raman}. 

\section{Effect of BaO$_{12}$ on polarizability of BTO}
In order to understand the contribution of BaO$_{12}$ polyhedra to the change in the polarizability and the dielectric constant, we have considered cubic centrosymmetric structure of BTO (5-atom cell) and calculated the dielectric constant for different displacements of Ba and Ti atoms along $x$-axis. Fig. S\ref{Ba_atom_neglect}(a) represents the evolution of $\epsilon_{xx}$ and $\epsilon_{yy}$ as a function of Ti displacement from its centrosymmetric position denoted by $d_{Ti}$ for two positions of Ba ($d_{Ba}$ = 0.00 and 0.08 \AA, where $d_{Ba}$ is the displacement of the Ba atom from its centrosymmetric position). Nearly parabolic dependence of $\epsilon_{xx}$ and $\epsilon_{yy}$ on $d_{Ti}$ is observed. This indicates the importance of the TiO$_{6}$ octahedra contribution to the polarizability of BTO. It is also observed that the two curves for different values of $d_{Ba}$ are nearly degenerate, indicating the lower  importance of the BaO$_{12}$ contribution to the polarizability of BTO. In Fig. S\ref{Ba_atom_neglect}(b), we plotted the variation of $\epsilon_{xx}$ and $\epsilon_{yy}$ as a function of $d_{Ba}$ for two values of $d_{Ti}$. As expected, the change in the polarizability is much smaller in comparison to that for the Ti displacement. Therefore, we have neglected the effect of BaO$_{12}$ octahedra from the model of the polarizability for BTO.  

\bibliographystyle{apsrev4-1} 
\providecommand{\noopsort}[1]{}\providecommand{\singleletter}[1]{#1}%